\documentclass[prx,aps,amsmath,amssymb,amsfonts,twocolumn,nofootinbib,longbibliography,superscriptaddress]{revtex4-1}
\usepackage{amssymb}
\usepackage{bm,mathrsfs}
\usepackage{graphicx}
\usepackage{epsfig}
\usepackage{amsmath,bbm}
\usepackage{amsfonts,amssymb}
\usepackage{times}
\usepackage{verbatim}
\usepackage[sort&compress]{natbib}
\usepackage{amsmath}
\usepackage{bm}
\usepackage{float}

\allowdisplaybreaks[4]
\usepackage[colorlinks,breaklinks,linkcolor=blue,anchorcolor=blue,citecolor=blue,urlcolor=blue]{hyperref}
\begin{document}

\title{High-fidelity interconversion between Greenberger-Horne-Zeilinger and $W$ states through Floquet-Lindblad engineering in Rydberg atom arrays }

\author{X. Q. Shao}
\email{shaoxq644@nenu.edu.cn}
\affiliation{Center for Quantum Sciences and School of Physics, Northeast Normal University, Changchun 130024, China}
\affiliation{Center for Advanced Optoelectronic Functional Materials Research, and Key Laboratory for UV Light-Emitting Materials and Technology
of Ministry of Education, Northeast Normal University, Changchun 130024, China}

\author{F. Liu}
\affiliation{Center for Quantum Sciences and School of Physics, Northeast Normal University, Changchun 130024,  China}

\author{X. W. Xue}
\affiliation{Center for Quantum Sciences and School of Physics, Northeast Normal University, Changchun 130024, China}

\author{W. L. Mu}
\affiliation{Center for Quantum Sciences and School of Physics, Northeast Normal University, Changchun 130024, China}

\author{Weibin Li}
\email{weibin.li@nottingham.ac.uk}
\affiliation{School of Physics and Astronomy, and Centre for the Mathematics and Theoretical Physics of Quantum Non-equilibrium Systems, The University of Nottingham, Nottingham NG7 2RD, United Kingdom}

\begin{abstract}
Greenberger-Horne-Zeilinger and $W$ states feature genuine tripartite entanglement that cannot be converted into each other by local operations and classical communication.
Here, we present a dissipative protocol for deterministic interconversion between Greenberger-Horne-Zeilinger and $W$ states of three neutral $^{87}$Rb atoms arranged in an equilateral triangle of a
two-dimensional array.
With three atomic levels and diagonal van der Waals interactions of Rydberg atoms, the interconversion between tripartite entangled states can be efficiently accomplished in the Floquet-Lindblad framework through the periodic optical pump and dissipation engineering. We evaluate the feasibility of the existing methodology using the experimental parameters accessible to current neutral-atom platforms. We find that our scheme is robust against typical noises, such as laser phase noise and geometric imperfections of the atom array. In addition, our scheme can integrate the Gaussian soft quantum control technique, which further reduces the overall conversion time and increases the resilience to timing errors and interatomic distance fluctuations.  The high-fidelity and robust tripartite entanglement interconversion protocol provides a route to save physical resources and enhance the computational efficiency of quantum networks formed by neutral-atom arrays.
\end{abstract}

\maketitle
\section{Introduction}
Quantum information processing relies crucially on multiparticle entanglement, an essential feature of quantum physics \cite{1,2,3,4, PhysRevA.68.042317, PhysRevA.69.052319, PhysRevLett.123.070505, PhysRevLett.125.230501, PhysRevX.9.021046, RevModPhys.92.025002}.  It has been demonstrated that  Greenberger-Horne-Zeilinger (GHZ) state and $W$ state belong to different entanglement classes, and hence are not equivalent under local operations assisted by classical
communication \cite{8,32}. Any single qubit loss in a three-particle GHZ state brings the other two to a mixed state with just classical correlation. In contrast, this does not completely damage the  pairwise entanglement in a three-particle $W$ state,
in which the remaining pair of qubits retain the greatest possible amount of entanglement compared to any other three-qubit state.
 Such fundamentally different characteristics can be utilized to solve distinct tasks in quantum information processing. For instance, GHZ state can be employed in quantum sensing to bring the sensitivity down to the Heisenberg limit \cite{leibfried2004toward, RevModPhys.89.035002, PhysRevLett.112.080801, reiter2017dissipative, PhysRevA.97.042337}, while $W$ state is a preferred candidate of ensemble-based quantum memory since they are more robust than GHZ states upon loss of a particle \cite{PhysRevA.65.022314, PhysRevA.72.022327, PhysRevLett.93.230501, PhysRevA.82.052308}.
In terms of quantum resources, these two entangled states have different computation power. D'Hondt and Panangaden have shown that, in the context of paradigmatic problems from classical distributed computing, the GHZ state is the only pure state capable of solving the distributed consensus problem without the need for classical post-processing, and the $W$ state is the only pure state capable of solving the leader election problem in anonymous quantum networks precisely  \cite{10.5555/2011665.2011668}. Due to the potential applications, GHZ state and $W$ state of three or more qubits have attracted much attention in both theory and experiment \cite{PhysRevLett.87.230404, neeley2010generation,grafe2014chip, PhysRevLett.117.040501, omran2019generation, PhysRevA.99.032323, PhysRevLett.125.203603, PhysRevLett.123.070508}.
A major issue of concern is the direct interconversion between them, which improves the computing efficiency of future quantum networks~\cite{https://doi.org/10.1002/qute.201900015,ILLIANO2022109092}.
 The existing inventive techniques, achieved in linear optical systems \cite{13,15}, cavity quantum electrodynamics systems \cite{11,10}, nuclear
magnetic resonance systems \cite{36}, and spin-chain systems \cite{68,12}, are either probabilistic, irreversible or require complicated interactions. This motivates us to search for alternative solutions to realize a high-fidelity state interconversion protocol.

Neutral-atom systems have become a promising platform for quantum computing and quantum simulation \cite{70,72,73,74,21,75, PhysRevLett.123.170503, PhysRevA.105.032417,
PhysRevApplied.18.044042, PhysRevX.12.021049, PhysRevLett.128.120503,71}, due to their inherent long lifetime in electronic ground states and strong dispersive two-body interactions in Rydberg states. On another hand,  experimental techniques have rapidly developed to arrange a large number of atoms in defect-free and programmable geometry in various dimensions while retaining individual addressability~\cite{doi:10.1126/science.aah3778, 10.1038/nature24622, 10.1038/s41586-021-03582-4, scholl2021quantum, doi:10.1126/science.abi8794,ebadi2022quantum,graham2022multi}.
Based on neutral atoms, deterministic interconversions between tripartite GHZ and $W$ states have been explored theoretically~\cite{16,17,69,67}, in which logical qubits are encoded in the hyperfine ground state and electronically excited Rydberg state.
A common feature of these schemes is that the interconversion operation is beyond the Rydberg blockade regime (i.e. via facilitation). Furthermore, it requires time-modulated pulses of different optical frequencies and Rabi frequencies to couple states among the whole Hilbert space of the three-particle symmetric states.
As a result, several practical limitations need to be addressed in the aforementioned protocols.
First, the facilitation condition (Rydberg antiblockade) demands precise control of the laser detuning to compensate for interactions in Rydberg levels. This poses experimental challenges as geometry imperfection of atom arrays cannot be ruled out completely \cite{PhysRevLett.98.023002, PhysRevLett.104.013001, 77, PhysRevLett.128.013603}. Even tiny variations of interatomic spacing cause large energy shifts~\cite{PhysRevLett.110.213005}, invalidate the facilitation condition. Second, polychromatic driving fields are needed to precisely control the Rabi frequencies. Though crucially important, such time-dependent coupling increases experimental difficulties \cite{PhysRevA.72.022347, PhysRevA.85.042310, PhysRevA.103.022424, PhysRevLett.124.070503}. Third, since the logic qubit is encoded in the Rydberg state, atomic spontaneous emission will inevitably play roles during the interconversion, and hence decreases the fidelity~\cite{PhysRevLett.111.033606, PhysRevLett.125.143601, PhysRevLett.124.043402, PhysRevResearch.4.L032046}. Moreover, the purity of the entangled state cannot be ensured if the dynamics are no longer unitary.

In response to the aforementioned difficulties, we propose a dissipative interconversion scheme between the tripartite GHZ and $W$ states using Rydberg atoms arranged in triangular arrays (Fig.~\ref{pulse}).
Based on $\Lambda$-type atomic levels, the dynamic evolution of the system is characterized by a Markovian master equation, where the coherent evolution is provided by the laser-atom interaction and the dissipation is provided by the controlled spontaneous emission.
Employing the periodic pump and dissipation engineering~\cite{kienzler2015quantum, PhysRevLett.128.080503,Mu:22}, we achieve deterministic interconversion of the above-entangled states in the
Floquet-Lindblad framework by altering the order of coherent driving fields \cite{PhysRevLett.117.250401, PhysRevLett.120.216801,ikeda2020general, PhysRevB.101.100301, PhysRevA.105.012208}.
Our protocol offers the following benefits: (i) The interacting Rydberg states are virtually excited during the Rydberg pump process. Unlike Rydberg antiblockade-based protocols \cite{16,17,69,67, PhysRevLett.125.133602, PhysRevLett.128.013603}, our approach is robust against variations in atomic location. (ii) We only need two intensities and frequencies of the laser to drive each pair of ground and excited states, which reduces the demand for laser power and renders the laser insensitive to fluctuations in intensity. (iii) The essential distinction between our method and previous ones is that we use spontaneous emission as a resource, allowing us to encode the qubits in the atomic ground states. This loosens the restriction on initial state purity.
The above advantages ensure that we can accomplish a high-fidelity and robust interconversion for two different kinds of genuine tripartite entanglement, which may save physical resources and boost the computing efficiency of quantum networks realized with neutral atoms.

The structure of this paper is organized as follows.
We present both the coherent and the incoherent operation steps in Sec.~\ref{II},  which are essential components in the construction of a quantum state interconversion mechanism. In Sec.~\ref{III}, we establish the scheme for the conversion from the GHZ state to the $W$ state, as well as the reverse procedure, for three Rydberg atoms arranged in an equilateral triangle.  The performance of the scheme is illustrated through numerical simulation.
Within the Floquet-Lindblad framework, we validate the uniqueness of each process's steady state.
Sec.~\ref{IV} delves into the experimental feasibility, including robustness to laser phase noise, the influence of distance fluctuations, and timing errors. We also compare our system to earlier schemes based on unitary evolutions. We conclude our work in Sec.~\ref{V}.

\section{controllable coherent and dissipative operations in Rydberg atom arrays}\label{II}
\begin{figure}
\centering\scalebox{0.21}{\includegraphics{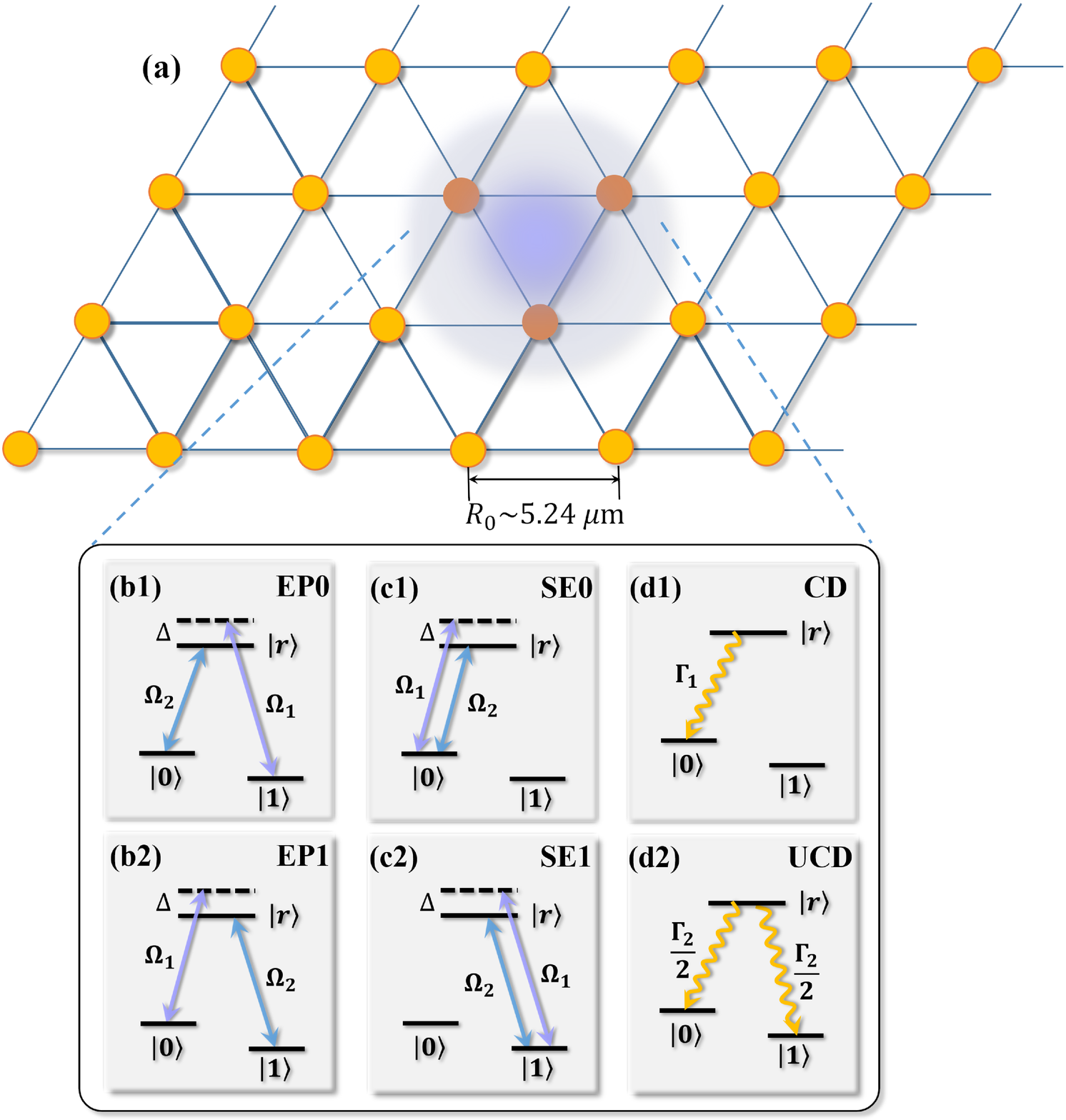}}
\caption{\label{pulse} Controllable coherent and incoherent operations in a neutral-atom triangular array. The $\Lambda$-type three-level atom consists of two hyperfine ground states  $|0\rangle$ and $|1\rangle$, and a Rydberg state $|r\rangle$.
The laser-atom interaction models in (b) and (c) realize the required entangled pump and selective excitation, respectively, and (d) depicts two distinct engineered spontaneous emission processes.}
\end{figure}

The system under investigation consists of three neutral $^{87}$Rb atoms arranged in an equilateral triangle, embedded in a larger two-dimensional array, as shown in Fig.~\ref{pulse}(a). Such two-dimensional arrays, with tens to hundreds of atoms, have been experimentally realized recently. Each atom can be individually controlled, whose spacing can be adjusted from $3$ to $10$~$\mu$$\rm m$ \cite{doi:10.1126/science.aah3778,10.1038/nature24622,10.1038/s41586-021-03582-4,scholl2021quantum,doi:10.1126/science.abi8794,ebadi2022quantum,graham2022multi}.
We consider a $\Lambda$ transition which consists of a Rydberg state $|r\rangle$, and two hyperfine ground states given by $|0\rangle$ and $|1\rangle$, as depicted in Fig.~\ref{pulse}.
In Rydberg states, atoms interact strongly via distance-dependent van der Waals (vdW) interaction~\cite{21}.

The interconversion scheme involves four coherent operations and two dissipative operations. All can be realized and controlled by laser-atom couplings. The configurations of the coherent operations are depicted in Figs.~\ref{pulse}(b) and \ref{pulse}(c), whereas Fig.~\ref{pulse}(d) illustrates the dissipation processes. They form the fundamental building blocks of quantum state interconversion. The function and realization of the above operations will be discussed below.

\subsection{Coherent operations}
In the coherent operation shown in Fig.~\ref{pulse}(b1),  two ground states $|0\rangle$ and $|1\rangle$ are coupled to Rydberg state $|r\rangle$ by a resonant (Rabi frequency  $\Omega_{2}$) and an off-resonant laser (Rabi frequency $\Omega_{1}$ and detuning $-\Delta$), respectively.
More details about the lasers can be found in
Appendix \ref{A}.
 In the interaction picture, the corresponding Hamiltonian reads $(\hbar=1)$
\begin{eqnarray}\label{fullmurp00}
H_{(b1)}&=&\sum_{j=1}^{3}\frac{\Omega_{1}}{2}e^{-i\Delta t}|r_j\rangle\langle 1_j|+\frac{\Omega_{2}}{2}|r_j\rangle\langle 0_j|+{\rm H.c.}\nonumber\\&&+\sum_{j<k}U_{rr}|r_j\rangle\langle r_j|\otimes|r_k\rangle\langle r_k|,
\end{eqnarray}
where  $U_{rr}=C_6/R^6$ is the vdW interaction with $C_6$ and $R$ being the dispersion coefficient and  interatomic distance.
In a strongly interacting regime provided $U_{rr}=\Delta\gg\Omega_1\gg\Omega_2$, multiple Rydberg excitation can be adiabatically eliminated, and the dynamics of the system are restricted to the ground state and a single Rydberg excitation sector. This yields an effective Hamiltonian of the three atoms,
\begin{eqnarray}\label{murp00}
H_{\textrm{EP0}}&=&\frac{\sqrt{3}\Omega_{2}}{2}|D_{0}\rangle\langle 000|-\frac{\sqrt{3}\Omega_{2}}{4}(|\psi_{0}^{2}\rangle+|\psi_{0}^{3}\rangle)\langle W_{0}^{\prime}|\nonumber\\&&-\frac{\Omega_{2}}{4}(2|\psi_{0}^{1}\rangle+|\psi_{0}^{2}\rangle-|\psi_{0}^{3}\rangle)\langle W_{0}^{\prime\prime}|+\textrm{H.c.},
\end{eqnarray}
where we have defined superposition states, $|D_{0}\rangle=(|r00\rangle+|0r0\rangle+|00r\rangle)/\sqrt{3},|\psi_{0}^{1}\rangle=(|0r1\rangle-|01r\rangle)/\sqrt{2},|\psi_{0}^{2}\rangle=(|r01\rangle-|10r\rangle)/\sqrt{2}, |\psi_{0}^{3}\rangle=(|r10\rangle-|1r0\rangle)/\sqrt{2}, |W_{0}^{\prime}\rangle=(2|100\rangle-|010\rangle-|001\rangle)/\sqrt{6}$, and $|W_{0}^{\prime\prime}\rangle=(|010\rangle-|001\rangle)/\sqrt{2}$ with  $|ijk\rangle=|i_1\rangle\otimes|j_2\rangle\otimes|k_3\rangle$ standing for three-atom Fock state. The derivation of the effective Hamiltonian can be found in Appendix \ref{B}.
Here, emphasis should be placed on two points. First,
due to the presence of the Rydberg blockade effect, the system can be excited to a maximum of a single excitation subspace. Second, we employ a mechanism similar to electromagnetically induced transparency (EIT) so that in addition to states containing two or more atoms in $|1\rangle$ being unaffected, there exists a peculiar entangled state $|W_{0}\rangle=(|100\rangle+|010\rangle+|001\rangle)/\sqrt{3}$ which is also the dark state of Eq.~(\ref{murp00}). Since Eq.~(\ref{murp00}) can pump the separable ground states through the $|0\rangle\rightarrow|r\rangle$ transition to the entangled states, we refer to this dynamic process as entangled pump 0 (EP0) for convenience of subsequent discussion.

Now if we exchange the parameters of the two laser fields driving the atom from the ground states to the Rydberg state, we can obtain the model of Fig.~\ref{pulse}(b2) and its effective Hamiltonian
\begin{eqnarray}\label{murp10}
H_{\textrm{EP1}}&=&\frac{\sqrt{3}\Omega_{2}}{2}|D_{1}\rangle\langle 111|-\frac{\sqrt{3}\Omega_{2}}{4}(|\psi_{1}^{2}\rangle+|\psi_{1}^{3}\rangle)\langle W_{1}^{\prime}|\nonumber\\&&-\frac{\Omega_{2}}{4}(2|\psi_{1}^{1}\rangle+|\psi_{1}^{2}\rangle-|\psi_{1}^{3}\rangle)\langle W_{1}^{\prime\prime}|+\textrm{H.c.},
\end{eqnarray}
with $|D_{1}\rangle=(|r11\rangle+|1r1\rangle+|11r\rangle)/\sqrt{3}, |\psi_{1}^{1}\rangle=(|1r0\rangle-|10r\rangle)/\sqrt{2}, |\psi_{1}^{2}\rangle=(|r10\rangle-|01r\rangle)/\sqrt{2}, |\psi_{1}^{3}\rangle=(|r01\rangle-|0r1\rangle)/\sqrt{2}, |W_{1}^{\prime}\rangle=(2|011\rangle-|101\rangle-|110\rangle)/\sqrt{6}$, and $ |W_{1}^{\prime\prime}\rangle=(|101\rangle-|110\rangle)/\sqrt{2}$. By this, the temporal evolution of the state $|W_{1}\rangle=(|110\rangle+|101\rangle+|011\rangle)/\sqrt{3}$ is frozen. Correspondingly, we call this dynamic process entangled pump 1 (EP1).

In Fig.~\ref{pulse}(c1), we employ two driving laser fields simultaneously to couple the transition between state $|0\rangle$ and Rydberg state $|r\rangle$.
The Hamiltonian in the interaction picture is provided by
\begin{eqnarray}\label{IIe10}
H_{(c1)}&=&\sum_{j=1}^{3}\frac{\Omega_{1}}{2}e^{-i\Delta t}|r_j\rangle\langle 0_j|+\frac{\Omega_{2}}{2}|r_j\rangle\langle 0_j|+{\rm H.c.}\nonumber\\&&+\sum_{j<k}U_{rr}|r_j\rangle\langle r_j|\otimes|r_k\rangle\langle r_k|.
\end{eqnarray}
By using a similar process for derivation from Eq.~(\ref{fullmurp00}) to Eq.~(\ref{murp00}),
the effective Hamiltonian under the condition $U_{rr}=\Delta\gg\Omega_1\gg\Omega_2$ is obtained as (see Appendix \ref{B})
\begin{equation}\label{urp00}
H_{\textrm{SE0}}=\frac{\Omega_{2}}{2}(|r11\rangle\langle 011|+|1r1\rangle\langle 101|+|11r\rangle\langle 110|)+\mathrm{H.c.}
\end{equation}
In contrast to Eq.~(\ref{murp00}), the action of Eq.~(\ref{urp00}) will only excite the atom in the separable ground state to the associated single excited state through the $|0\rangle\rightarrow|r\rangle$ transition when all other atoms are in the uncoupled state $|1\rangle$. Consequently, we refer to this process as selective excitation 0 (SE0).

Similarly, we can apply two lasers to the transitions between state $|1\rangle$ and Rydberg state $|r\rangle$ under identical circumstances to achieve the following selective excitation 1 (SE1) Hamiltonian
\begin{equation}\label{urp10}
H_{\textrm{SE1}}=\frac{\Omega_{2}}{2}(|r00\rangle\langle 100|+|0r0\rangle\langle 010|+|00r\rangle\langle 001|)+\mathrm{H.c.}
\end{equation}

In the above equations, we have made certain approximations. The accuracy of the effective Hamiltonian has been validated by comparing the dynamics of the full Hamiltonian, which is shown in Appendix \ref{B}. These simulations show an excellent agreement with the results obtained by the effective Hamiltonian.

\subsection{Dissipative operations}
\begin{figure*}
\centering\scalebox{0.35}{\includegraphics{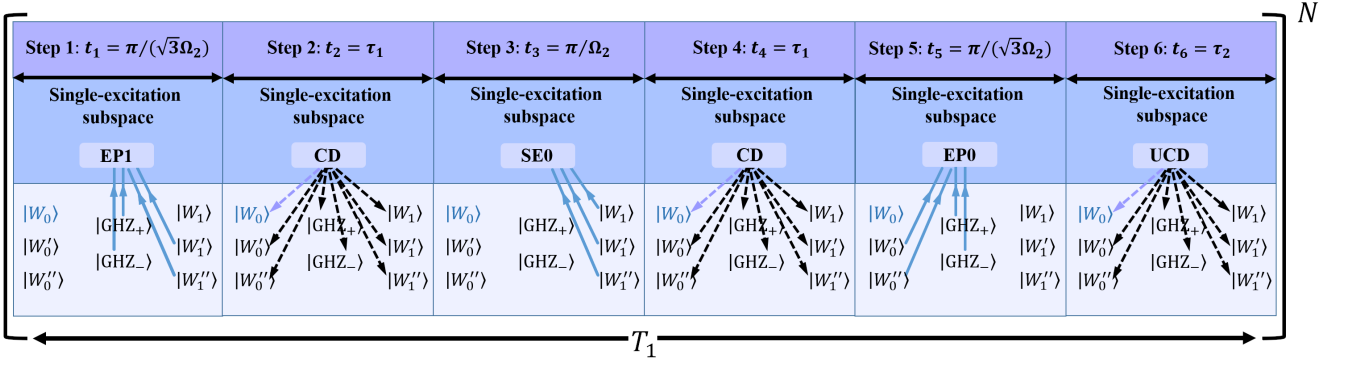}}
\caption{\label{W3}The steps for the scheme of conversion $\mathcal{I}$ in each cycle.
 The top portion of the figure depicts the timing of each step, while the bottom portion depicts how each step influences the states of the system.
 In steps $(1)$ and $(5)$, we utilize the entangled pump to destroy the populations of $|000\rangle$ and $|111\rangle$ by selecting the time $t_{1}=t_{5}=\pi/(\sqrt{3}\Omega_{2})$, while leaving the populations of $|W_1\rangle$ and $|W_0\rangle$ unchanged, respectively. Step $(3)$ implements the selective excitation for state $|0\rangle$ to destabilize the steady state $|W_1\rangle$ of step $(1)$. Steps $(2)$, $(4)$, and $(6)$ are the engineered dissipative processes, with steps $(2)$, $(4)$ utilizing CD and step $(6)$ using UCD.}
\end{figure*}
Here we propose a quicker convergence approach to the desired state using a larger relaxation rate. To fulfill the demands of our scheme and speed up the spontaneous emission
of the Rydberg state, we build two engineered spontaneous
emission channels, as illustrated in Figs.~\ref{pulse}(d1) and \ref{pulse}(d2), respectively. The corresponding dynamics are characterized by the master equations
\begin{equation}\label{dd1}
\dot{\rho}=\Gamma_{1}\mathcal{D}[|0\rangle\langle r|]\rho,
\end{equation}
and
\begin{equation}\label{dd2}
\dot{\rho}=\frac{\Gamma_{2}}{2}(\mathcal{D}[|0\rangle\langle r|]\rho+\mathcal{D}[|1\rangle\langle r|]\rho),
\end{equation}
where $\mathcal{D}[\bullet]\rho=\bullet\rho\bullet^{\dag}-(\bullet^{\dag}\bullet\rho+
\rho\bullet^{\dag}\bullet)/2$, and
 the effective decay rate $\Gamma_{j}=\Omega_{d_{j}}^{2}/\mathrm{\gamma}_j$ ($j=1,2$) with $\gamma_1^{-1}$ ($\gamma_2^{-1}$) being the lifetime of $5P_{3/2}$ ($5P_{1/2}$) and $\Omega_{d_{1}}$ ($\Omega_{d_{2}}$) the Rabi frequency resonantly coupled to the transition between the intermediate state $5P_{3/2}$ ($5P_{1/2}$) and Rydberg state $|r\rangle$.
 Both effective decays can be adjusted to a rate significantly greater than the spontaneous emission rate $\gamma_{r}$ of the Rydberg state (see Appendix \ref{C}).
Note that in Eq.~(\ref{dd2}),  the engineered decay rates of  $|r\rangle$ to $|0\rangle$ and $|1\rangle$ are considered to be equal for simplicity. In general, the branching ratio will vary depending on the specific atomic energy levels selected,  however, it does not affect the scheme's performance. Based on the different decay modes of the system,
 we here divide these two processes into conditional decay (CD) and unconditional decay (UCD). Unlike the earlier demonstrations of dissipative protocols with continuous driving fields \cite{PhysRevLett.111.033607, PhysRevA.89.052313, PhysRevA.92.022328, PhysRevA.101.042328}, our scheme does not rely on a delicate balance between unitary and incoherent evolution, and leads to a faster decay process.

\section{Dissipative Interconversion between GHZ state and $W$ state}\label{III}
\subsection{GHZ-to-$W$ state conversion $\mathcal{I}$}\label{III0}

 \begin{figure}
\centering\scalebox{0.46}{\includegraphics{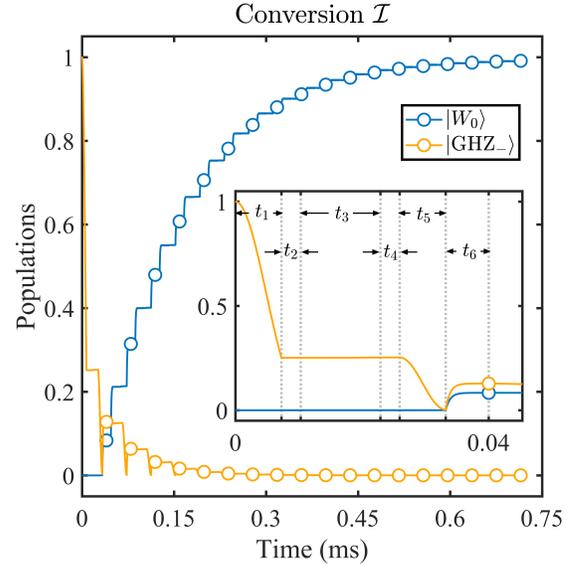}}
\caption{\label{GHZW}The temporal evolution of populations for states $|\textrm{GHZ}_{-}\rangle$ and $|W_{0}\rangle$ during the conversion process $\mathcal{I}$. The parameters are selected as $\Omega_{1}=2\pi\times4~\rm MHz$, $\Omega_{2}=2\pi\times0.04~\rm MHz$,  $U_{rr}=\Delta=2\pi\times200~\rm MHz$, $\Gamma_{1}=2\pi\times0.2424~\rm MHz$, $\Gamma_{2}=2\pi\times0.23~\rm MHz$,  $\gamma_{r}=2\pi\times0.28~\rm kHz$, $\tau_{1}\simeq3~\mu\mathrm{s}$, and $\tau_{2}\simeq6.7~\mu\mathrm{s}$. The corresponding relaxation time is about $0.715$~$\rm ms$.
The inset depicts the population changes in the two states throughout the first cycle.}
\end{figure}
This section describes the dissipation-based one-way conversion of $\mathcal{I}$ from the state $|\mathrm{GHZ}_{-}\rangle=(|000\rangle-|111\rangle)/\sqrt{2}$  to state $|W_0\rangle$. To demonstrate the benefits of the strategy, we choose a generic mixed state in the ground state subspace rather than the pure GHZ state as the initial state to analyze the specific steps of the scheme. As illustrated in Fig.~\ref{W3}, the entire conversion $\mathcal{I}$ consists of multiple cycles, each containing six steps.

Step $(1)$: We begin with a laser-cooling-prepared initial state of the form $\rho(0)=a_1|W_0\rangle\langle W_0|+a_2|W_0^{'}\rangle\langle W_0^{'}|+a_3|W_0^{''}\rangle\langle W_0^{''}|+a_4|W_1\rangle\langle W_1|+a_5|W_1^{'}\rangle\langle W_1^{'}|
+a_6|W_1^{''}\rangle\langle W_1^{''}|+a_7|{\rm GHZ_+}\rangle\langle {\rm GHZ_+}|+a_8|{\rm GHZ_-}\rangle\langle {\rm GHZ_-}|$, where $\sum_{i=1}^8 a_i=1$ and ${|\rm GHZ_+}\rangle=(|000\rangle+|111\rangle)/\sqrt{2}$.
Under the action of EP1 of Eq.~(\ref{murp10}), states ${|\rm GHZ_{\pm}}\rangle$ are excited to $(|000\rangle\pm|D_1\rangle)/\sqrt{2}$ at time $t_{1}=\pi/(\sqrt{3}\Omega_{2})$, while the populations of states $|W_1^{'}\rangle$ and $|W_1^{''}\rangle$ are partially converted to the corresponding single-excitation subspace $\{|\psi_{1}^{1}\rangle$, $|\psi_{1}^{2}\rangle$, $|\psi_{1}^{3}\rangle\}$.

Step $(2)$: We switch off the coherent driving EP1 and switch on the dissipation channel CD for some time $t_2=\tau_1$ until the Rydberg state $|r\rangle$ completely decays into state $|0\rangle$, and this process leads to a one-way increase in the population of state $|W_0\rangle$.

\begin{figure*}
\centering\scalebox{0.35}{\includegraphics{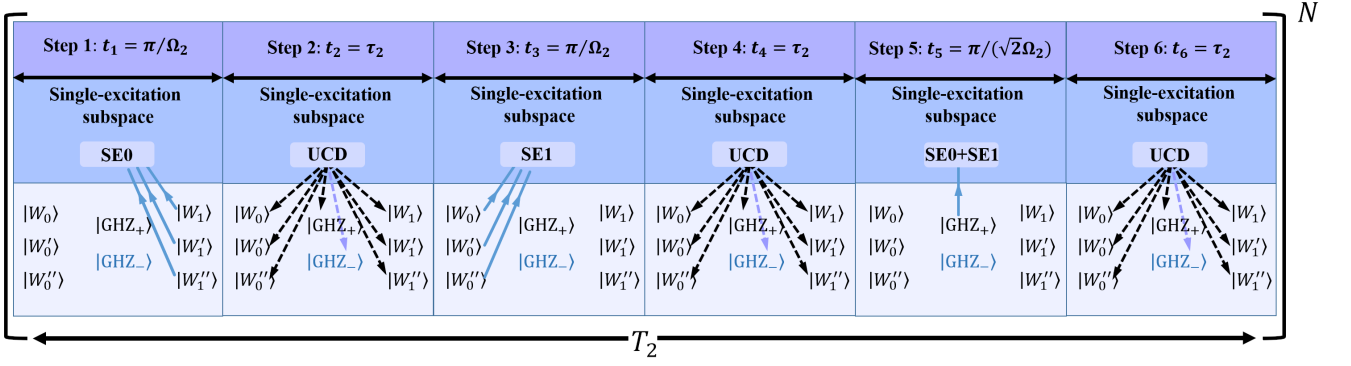}}
\caption{\label{GHZ3}The steps for the scheme of conversion $\mathcal{II}$ in each cycle. In steps $(1)$ and $(3)$, we employ the selective excitation to drive atoms into single-excitation subspace with time $t_{1}=t_{3}=\pi/\Omega_{2}$, respectively. Step $(5)$ employs selective excitation acting on state $|+\rangle$ to eradicate the population of $|\mathrm{GHZ}_{+}\rangle$ with time $t_{5}=\pi/(\sqrt{2}\Omega_{2})$. Steps $(2)$, $(4)$, and $(6)$ are the engineered dissipative processes constructed by the UCD.}
\end{figure*}

Step $(3)$: To  destabilize the population of state $|W_1\rangle$ in step $(1)$ while keeping the target state $|W_0\rangle$ constant, we employ the SE0 of Eq.~(\ref{urp00}). A time of $t_{3}=\pi/\Omega_{2}$ causes the states $|W_1\rangle$, $|W^{'}_1\rangle$, and $|W^{''}_1\rangle$ to transform into $|D_1\rangle$,
$(2|r11\rangle-|1r1\rangle-|11r\rangle)/\sqrt{6}$, and $ (|1r1\rangle-|11r\rangle)/\sqrt{2}$, respectively.

Step $(4)$: We turn off the coherent driving SE0 and restart the dissipation channel CD for a period $t_4=\tau_1$, and we see a further rise in the population of the target state $|W_0\rangle$.

Step $(5)$: Similar to step $(1)$, we implement EP0 of Eq.~(\ref{murp00}) to pump atoms from the ground states ${|\rm GHZ_{\pm}}\rangle$ upwards to $(|D_0\rangle\pm|111\rangle)/\sqrt{2}$ as $t_5=\pi/(\sqrt{3}\Omega_{2})$. Correspondingly, other states $|W_0^{'}\rangle$ and $|W_0^{''}\rangle$ will superimpose with the single-excitation subspace $\{|\psi_{0,1}^{0}\rangle$, $|\psi_{0,2}^{0}\rangle$, $|\psi_{0,3}^{0}\rangle\}$.

Step $(6)$: We deactivate the coherent driving EP0 and activate the engineered dissipation UCD for a duration $t_6=\tau_2$ that guarantees the Rydberg state will completely decay into the ground states. As a result, the population of the state $|W_0\rangle$ grows once again.

To make concrete simulations that are relevant to current experiments,  we consider the Rydberg state $|r\rangle\equiv|80S_{1/2},m_{J}=1/2\rangle$, and two hyperfine ground states $|0\rangle\equiv|5S_{1/2}, F=2,m_{F}=2\rangle$ and $|1\rangle\equiv|5S_{1/2}, F=1,m_{F}=1\rangle$ as the relevant levels depicted in Fig.~\ref{pulse}.
The separation between atoms is set to be greater than 5~$\mu$$\rm m$ since the dispersion coefficient $C_{6}/(2\pi)=4161.55$ ${\rm GHz}$ $\cdot~(\mu \mathrm{m})^{6}$ of the vdW interaction is highly compatible with the non-perturbative calculations \cite{78}.
After $18$ cycles, the system that was originally in state $|\rm GHZ_-\rangle$ is converted to state $|W_0\rangle$ with a probability of more than $99\%$, as shown in  Fig.~\ref{GHZW}. In the simulation,  the spontaneous emission of the Rydberg state has been taken into account in each step, which barely affects the conversion efficiency.

\subsection{$W$-to-GHZ state conversion $\mathcal{II}$}\label{IV0}

\begin{figure}
\centering\scalebox{0.46}{\includegraphics{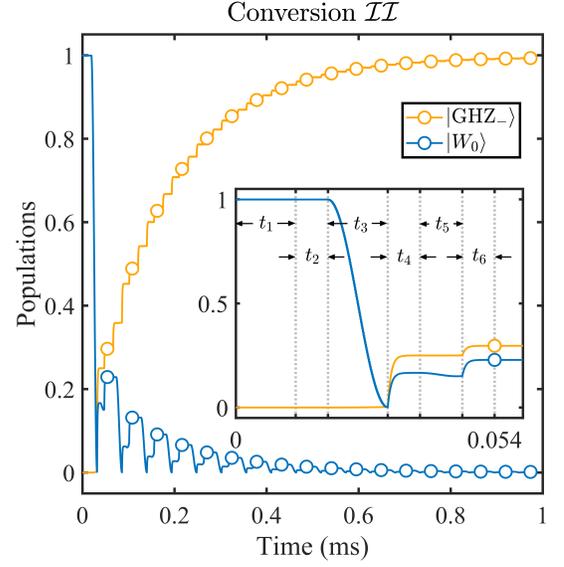}}
\caption{\label{WGHZ}
The temporal evolution of populations for states $|W_{0}\rangle$ and $|\textrm{GHZ}_{-}\rangle$ during the conversion process $\mathcal{II}$. The parameters are the same as in Fig.~\ref{GHZW}. The inset depicts the population variations in both states during the first cycle.}
\end{figure}

We now turn to examine the unidirectional conversion $\mathcal{II}$ from state $|W_0\rangle$ to state $|\rm GHZ_-\rangle$ dissipatively. Without loss of generality, we continue to utilize $\rho(0)=a_1|W_0\rangle\langle W_0|+a_2|W_0^{'}\rangle\langle W_0^{'}|+a_3|W_0^{''}\rangle\langle W_0^{''}|+a_4|W_1\rangle\langle W_1|+a_5|W_1^{'}\rangle\langle W_1^{'}|
+a_6|W_1^{''}\rangle\langle W_1^{''}|+a_7|{\rm GHZ_+}\rangle\langle {\rm GHZ_+}|+a_8|{\rm GHZ_-}\rangle\langle {\rm GHZ_-}|$ as the initial state of system.
Each cycle of the scheme, much like the conversion $\mathcal{I}$, consists of six steps, as depicted in Fig.~\ref{GHZ3}.

Step $(1)$: We achieve SE0 of Eq.~(\ref{urp00}) to drive the transition from the state $|0\rangle$ to the state $|r\rangle$, and then pump the states $|W_1\rangle$, $|W^{'}_1\rangle$, and $|W^{''}_1\rangle$ in the ground state subspace to $|D_1\rangle$,
$(2|r11\rangle-|1r1\rangle-|11r\rangle)/\sqrt{6}$, and $ (|1r1\rangle-|11r\rangle)/\sqrt{2}$ after $t_1=\pi/\Omega_2$, while all other states stay unaffected.

Step $(2)$: To raise the population of target state $|\rm GHZ_-\rangle$ in a unidirectional manner, we turn off the coherent driving SE0 and turn on the engineered dissipation UCD for a duration $t_2=\tau_2$ sufficient to allow the Rydberg state $|r\rangle$ to entirely decay into the ground states $|0\rangle$ and $|1\rangle$.

Step $(3)$: We employ SE1 of Eq.~(\ref{urp10}) to pump the states $|W_0\rangle$, $|W_0^{'}\rangle$, and $|W_0^{''}\rangle$ from the ground state subspace to states $|D_0\rangle$,
$(2|r00\rangle-|0r0\rangle-|00r\rangle)/\sqrt{6}$, and $ (|0r0\rangle-|00r\rangle)/\sqrt{2}$ with the same action time $t_3=\pi/\Omega_2$ as we did in step $(1)$.

Step $(4)$: When we resume the dissipation channel UCD for a time interval $t_4=\tau_2$ after switching off the coherent driving $H_{\rm SE1}$, the population of state $|\rm GHZ_-\rangle$ continues to grow.

Step $(5)$: The first four steps can guarantee a progressive rise in the population of the mixed-state system made up of
$|\rm GHZ_+\rangle=(|+++\rangle+|+--\rangle+|-+-\rangle+|--+\rangle)/2$ and
$|\rm GHZ_-\rangle=(|---\rangle+|++-\rangle+|+-+\rangle+|-++\rangle)/2$. To further purify the system,  we combine the coherent driving in steps $(1)$ and $(3)$ to simultaneously drive the transition from states $|0\rangle$ and $|1\rangle$ to state $|r\rangle$, thereby realizing the following form of Hamiltonian
$H_{\mathrm{SE+}}={\Omega_{2}}/{\sqrt{2}}(|r--\rangle\langle +--|+|-r-\rangle\langle -+-|+|--r\rangle\langle --+|)+\mathrm{H.c.}$. Governed by this Hamiltonian, some components (such as $|+--\rangle$, $|-+-\rangle$, and $|--+\rangle$) in state $|\rm GHZ_+\rangle$ may be fully pumped to the single-excitation subspace corresponding to the time $t_5=\pi/(\sqrt{2}\Omega_2)$, while the target state $|\rm GHZ_-\rangle$ remains stable.

Step $(6)$: We reactivate the engineered dissipation UCD for some time $t_6=\tau_2$ to redistribute the populations of ground states, and the population of $|\rm GHZ_-\rangle$ will be enhanced once more.

In Fig.~\ref{WGHZ}, the temporal evolution of populations for states $|W_{0}\rangle$ and $|\textrm{GHZ}_{-}\rangle$ during the conversion process $\mathcal{II}$ is shown. We find that after $18$ cycles $(\sim0.973~\rm ms)$, the system has arrived at the desired state $|\textrm{GHZ}_{-}\rangle$ with a probability that is greater than $99\%$.

Now we finish the interconversion between the GHZ  and $W$ states of Rydberg atoms based on periodically collective laser pump and
dissipation. The population of the target state progressively increases with each cycle, and the required time is significantly shorter than the continuous driving protocols from the perspective of dissipation.
More importantly, the employment of coherent pump and dissipation in alternating fashion removes the dependency on extra degrees of freedom (such as optical cavities and motional mode of trapped ions)  \cite{PhysRevA.96.062315, Li:18, PhysRevLett.117.040501}, considerably reducing the operational complexity of the experiment.

\subsection{Unique steady state  in the Floquet-Lindblad framework}

In the previous subsections, we introduced dissipative quantum state interconversion and demonstrated its feasibility through dynamic evolution. This part establishes that the target state of the related quantum state conversion process is the unique steady state using more rigorous mathematical and numerical methods.
With the help of Fock-Liouville space \cite{doi:10.1063/1.5115323}, we can map the density matrix $
{\rho}=\sum_{i,j=1}^N\rho_{i,j}|i\rangle\langle j|$  of the three-atom system defined in 27-dimensional Hilbert space into
\begin{equation}\label{f2}
|{\rho}\rangle\rangle=\sum_{i,j=1}^N\rho_{i,j}|i\rangle\otimes|j\rangle^*,
\end{equation}
and the time evolution of the system now corresponds to the
matrix equation
\begin{equation}\label{f4}
\frac{d|{{\rho}}\rangle\rangle}{dt}=\hat{\mathcal{L}}|{\rho}\rangle\rangle,
\end{equation}
where $\hat{\mathcal{L}}$ is the Liouvillian superoperator whose spectral properties determine whether there is a unique steady state in the dissipative system. For a time-independent $\hat{\mathcal{L}}$, it is easy to get the evolution of the system in the
form
\begin{equation}\label{f5}
|{\rho}(t)\rangle\rangle=e^{\hat{\mathcal{L}t}}|{\rho}(0)\rangle\rangle.
\end{equation}

\begin{figure}
\centering\scalebox{0.28}{\includegraphics{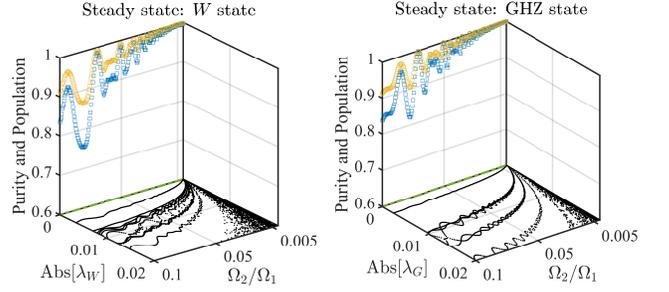}}
\caption{\label{liuw} The variations of purity $\mathcal{P}_{W(G)}$, population $P_{W(G)}$, and spectrum (in the unit of $2\pi\times\rm 1~MHz$) of the effective Liouvillian matrix $\hat{\mathcal{L}}^{W(G)}_{\rm eff}$ with $\Omega_2/\Omega_1$ for the dissipative preparation of $W$ state (left panel) and GHZ state (right panel). The parameters are chosen as $\Omega_1=2\pi\times4~\rm MHz$, $U_{rr}=\Delta=2\pi\times200~\rm MHz$, and other parameters are consistent with Figs.~\ref{GHZW} and \ref{WGHZ}, respectively. }
\end{figure}

Since our scheme uses periodic pump and dissipation, the effective Liouvillian superoperator for each cycle can be expressed  in logarithmic
form as
\begin{equation}\label{f6}
\hat{\mathcal{L}}^{W(G)}_{\rm eff}=\frac{1}{T}{\rm ln}(e^{\hat{\mathcal{L}}_6t_6}e^{\hat{\mathcal{L}}_5t_5}e^{\hat{\mathcal{L}}_4t_4}
e^{\hat{\mathcal{L}}_3t_3}e^{\hat{\mathcal{L}}_2t_2}e^{\hat{\mathcal{L}}_1t_1}),
\end{equation}
where the period $T=\sum_{i=1}^6t_i$ and the subscript indicate the step in each cycle for preparation of $W$ (GHZ) state.
By denoting the eigenvalues of $\hat{\mathcal{L}}^{W(G)}_{\rm eff}$ as $\lambda_{W(G)}$, we describe the variations of the modulus of  $\lambda_{W(G)}$ concerning $\Omega_2/\Omega_1$ in the $x-y$ plane of Fig.~\ref{liuw}, where there is always a zero eigenvalue (green dash-dotted line), which indicates that the dissipative system has a unique steady state.
Upon further investigation of the quantum state corresponding to the zero eigenvalue by specifying the purity
$
\mathcal{P}_{W(G)}= {\rm Tr}({\rho}^2)
$ (blue empty square)
and population
$
P_{W(G)} = \langle W_0({\rm GHZ}_-)|{\rho}|W_0({\rm GHZ}_-)\rangle
$ (yellow empty circle)
in the $x-z$ plane, it is discovered that when $\Omega_2/\Omega_1$ is within $0.025$, the system may be stabilized with high fidelity in the target state. This proves once again the accuracy of the preceding approximation criteria ($\Omega_1\gg\Omega_2$).

\section{Discussion of the experimental feasibility}\label{IV}
\subsection{Robustness against the laser phase noise}\label{VI00}
\begin{figure*}
\centering
\hspace*{-1.7cm}
\scalebox{0.46}{\includegraphics{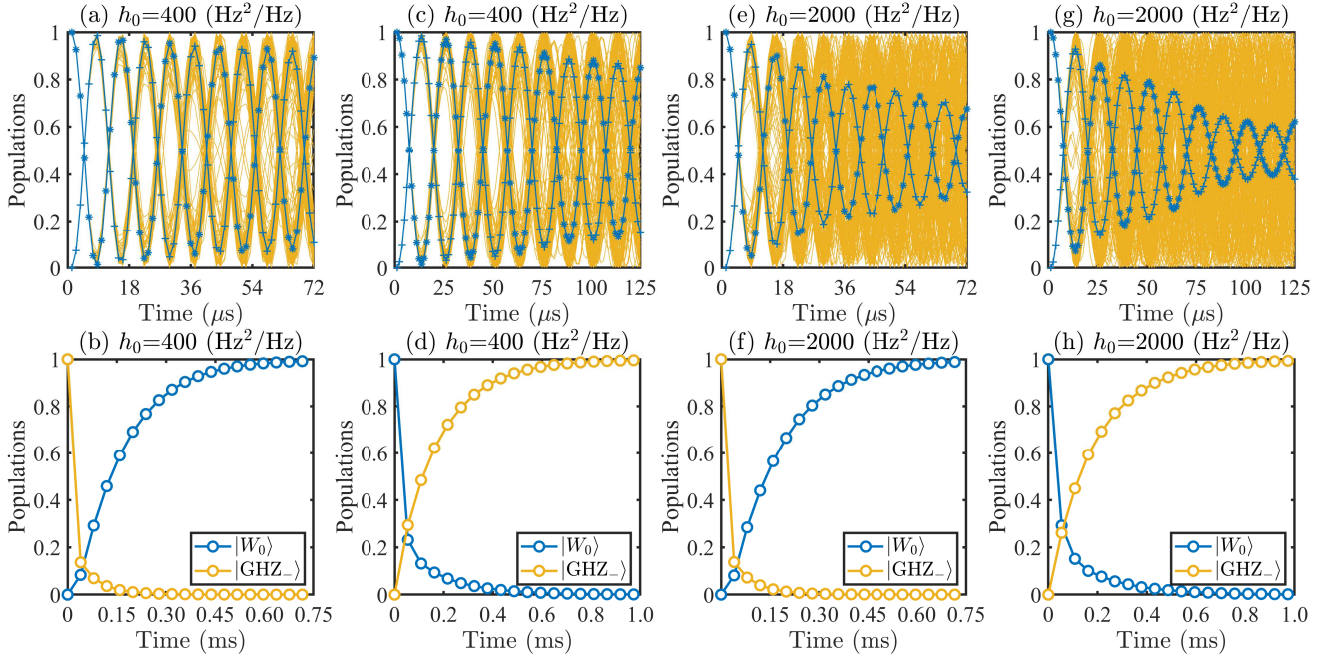}}
\caption{\label{noise}Influence of laser phase noise on the Rabi oscillation and quantum state interconversion. (a) and (c) illustrate many simulated Rabi oscillations and their average of some representative quantum states with a typical white-noise amplitude of $h_0=400~\rm Hz^2/Hz$.
(b) and (d) show the corresponding average of five realizations of quantum state conversions with the same noise parameter. The contents of (e) and (g) are similar to that of (a) and (c), with the exception that the noise amplitude is set to $h_0=2000~\rm Hz^2/Hz$. Similar to (b) and (d), (f) and (h) exhibit the same temporal evolution process at a noise level of $h_0=2000~\rm Hz^2/Hz$.}
\end{figure*}
The phase fluctuations of the classical field are the typical decoherence factors in Rydberg atom experiments, which have been effectively controlled in recent experiments \cite{PhysRevLett.121.123603,madjarov2020high, PhysRevA.105.042430}. In Ref.~\cite{PhysRevApplied.15.054020}, the  laser linewidths are optimized to be less than
$1~\rm kHz$ when the Rydberg-excitation lasers are frequency stabilized by
the Pound-Drever-Hall technique with a tunable reference
cavity and the phase noise of the excitation lasers is further suppressed
by nearly three orders below $1~\rm MHz$ by  employing the transmitted light from the high-finesse cavity to install an extra injection lock in the laser
diodes.
To accurately measure the influence of laser phase noise, it is mathematically necessary to obtain the power spectral density of noise or the related frequency spectral density \cite{PhysRevA.97.053803, PhysRevA.99.043404}. These quantities are closely related to specific experiments and are not suitable for general discussion.

In recent theoretical work, Jiang {\it et al}. developed a model to identify the characteristics observed in the laser self-heterodyne noise spectrum, and in the weak-noise regime, the analytical theory is in good agreement with the numerical simulation including the phase noise \cite{jiang2022sensitivity}. This study motivates us to use the time-series expansion of the laser noise
to examine the impact of laser noise on our proposal.
Since the resonant laser dominates the effective dynamics of the system, we are primarily concerned with the influence of the phase noise of the Rabi frequency $\Omega_2$. By introducing a random phase $\phi(t)$, the Rabi frequency can be modified as
\begin{equation}
\Omega_2\rightarrow\Omega_2e^{i\phi(t)},
\end{equation}
where the time traces of the laser phase fluctuations are defined as \cite{tucker1984numerical,merigaud2017free}
\begin{equation}
\phi(t)=\sum_{j=1}^{\infty}2\sqrt{S_{\phi}(f_j)\Delta f}\cos(2\pi f_jt+\varphi_j),
\end{equation}
where $S_{\phi}(f)$ is the laser phase power spectral density, $f_j=j\Delta f$, and
the random variables $\varphi_j$ are uniformly distributed over $[0, 2\pi]$.
For the numerical solution of the master equation by the Runge-Kutta method, the time traces of the laser frequency fluctuations are more useful, i.e.,
\begin{equation}
\delta v(t)=\frac{1}{2\pi}\frac{d\phi}{dt}=-\sum_{j=1}^{\infty}2\sqrt{S_{\delta v}(f_j)\Delta f}\sin(2\pi f_jt+\varphi_j),
\end{equation}
where $S_{\delta v}(f)=f^2S_{\phi}(f)$ is the laser frequency power spectral density. Here we suppose the error is dominated by the white-noise background with constant noise spectrum  $S_{\delta v}=h_0$,
 and a frequency bandwidth of $f_{M/2}=10~\rm MHz$ is large enough for our purpose, where $M/2=500$ is the number of discrete frequency components and $\Delta t=1/(M\Delta f)$ according to the Nyquist sampling theorem.
 In Figs.~\ref{noise}(a) and \ref{noise}(c), we first consider a typical white-noise amplitude ($h_0=400~\rm Hz^2/Hz$) and investigate the Rabi oscillation of some representative quantum states, such as states $|000\rangle$ and $|D_0\rangle$ governed by $H_{\rm EP0}$, and states $|110\rangle$ and $|11r\rangle$ governed by $H_{\rm SE0}$, respectively. The average results indicated by scattered points over 100 realizations (thin lines) display a slow damping of the oscillation. Correspondingly, Figs.~\ref{noise}(b) and \ref{noise}(d) show the average of five realizations of quantum state conversions under the same conditions.
 For step $5$ in the $W$-to-GHZ state conversion $\mathcal{II}$, we have assumed for convenience that both laser fields with Rabi frequency $\Omega_2$ have the same noise spectra.
 We see that at the end of the scheme (after 18 cycles), the populations of $W$ state and GHZ state can still reach $99.17\%$ and $99.56\%$, respectively.
The dephasing effect becomes more obvious with a bigger white-noise amplitude ($h_0=2000~\rm Hz^2/Hz$). As shown in Figs.~\ref{noise}(e) and \ref{noise}(g), the $1/e$ coherence times are both less than 5 Rabi cycles. Nevertheless, since our scheme alternately uses pump and dissipation, we only care about the efficiency of the first half of the Rabi cycle. Even if the quantum states do not completely flip in the unitary dynamics process, the subsequent dissipation and cyclic evolution can still make the population of the target state gradually accumulate. It can be seen from Figs.~\ref{noise}(f) and \ref{noise}(h) that after 18 cycles of evolution, the populations of $W$ state and GHZ state remain at $98.85\%$ and $99.31\%$, respectively. This demonstrates that our scheme is resistant to laser phase fluctuations.

\subsection{Effects of distance fluctuations and timing errors}\label{dt}
In the derivations of the entangled pump and selective excitation, we need an equation identical to the facilitation condition: $U_{rr}=\Delta$. This condition is difficult to strictly achieve in realistic physical systems, due to the thermal motion of atoms caused by finite temperature. Intuitively, a violation of this equation would have a detrimental impact on the scheme. Nevertheless, it should be noted that our approach employs this requirement to prevent the atoms from being pumped from a single-excited state to a double-excited state (analogous to the EIT process), so the double-excitation  Rydberg states are virtually excited. Within a given fluctuation range of atomic spacing, it is thus still possible to reach the desired state with great precision.
The atomic vibration also changes the optical intensity that the atoms experience, and when this is combined with the intensity fluctuations that are inherent to the laser fields themselves, the Rabi frequency used to characterize the coupling strength between the external field and the atom cannot remain constant. Therefore, for the coherent operations required by our scheme, which depends on the evolution time, the influence of the fluctuations of the Rabi frequency (particularly for $\Omega_2$) can be equivalently reflected as the errors in the time selection.

Refer to the experimental setup~\cite{labuhn2016tunable}, single $^{87}$Rb atoms at temperature of roughly $30~\mu\rm K$ are trapped in optical traps with a wavelength of $\lambda_{f}=850~\rm nm$ and a $1/e^2$ radius of $\omega_{f}=1~\mu\rm m$. For a power of about $3.45~\rm mW$, the trap has a typical depth $U_{f}=1~\rm mK$. Under these parameters, the trap frequencies are $\omega_{z}=2\pi\times18.99~\rm kHz$, $\omega_{x,y}=2\pi\times99.26~\rm kHz$, resulting in the position uncertainties of $\sigma_{z}=452.67~\rm nm$ and $\sigma_{x,y}=86.6~\rm nm$. The correlated vdW energy displacement was calculated in Ref.~\cite{PhysRevLett.118.063606} as $\overline{\delta U_{rr}}=6|U_{rr}(R_0)|\overline{\delta R}/R_0$, with ${\overline{\delta R}}\simeq\sqrt{2}\sigma_x$ for two neighboring atoms separated by an average distance $\mathbf{{R_0}}=(R_0,0,0)$.
For the vdW potential employed in the simulation, $U_{rr}=2\pi\times200~{\rm MHz}$, which corresponds to the interatomic distance $R_0=5.2445~\mu \rm m$, the energy displacement of the vdW is about $\overline{\delta U_{rr}}\simeq 2\pi\times28~\rm MHz$.
\begin{figure}
\centering
\scalebox{0.44}{\includegraphics{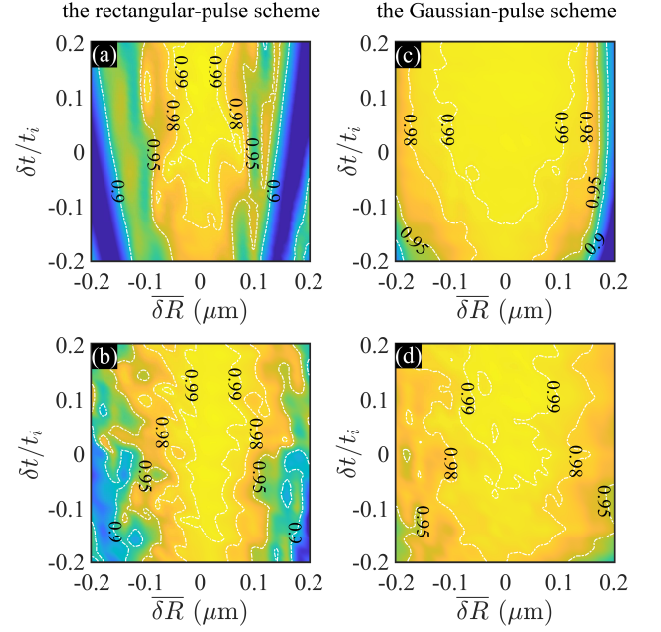}}
\caption{\label{error} (a) and (b) illustrate the effects of distance errors and timing errors by using the rectangular pulse, while (c) and (d) illustrate the same effects by utilizing the Gaussian pulse. (a) and (c) represent the GHZ-to-$W$ state conversion $\mathcal{I}$, whereas (b) and (d) represent the $W$-to-GHZ state conversion $\mathcal{II}$.}
\end{figure}
In Figs.~\ref{error}(a) and \ref{error}(b), we replicate the effect of distance variations on the conversions $\mathcal{I}$ and $\mathcal{II}$ across a greater range ${\overline{\delta R}}\in[-200~{\rm nm},200~{\rm nm}]$ and apply the errors of $\delta t\in[-0.2t_i,0.2t_i]$ to the action time of the rectangular pulse of Rabi frequency $\Omega_2$. Our approach has some resistance to these two variables. Due to the benefit of dissipative dynamics, the performance of the interconversion process may theoretically be further enhanced by increasing the number of cycles.

Alternatively, the robustness of the scheme may be improved by substituting rectangular pulses with Gaussian pulses without changing the number of cycles. This temporal modulation significantly suppresses the non-resonant contribution of the interaction while remaining insensitive to the operation time~\cite{66, Yin:21}. Now we change the time-independent Rabi frequency $\Omega_2$ into a time-dependent Gaussian form
\begin{equation}
\Omega_2(t)=\Omega_0\mathrm{exp}[-\frac{(t-2\sigma)^{2}}{2\sigma^{2}}],
\end{equation}
where $\Omega_0$ is the maximum amplitude and
the full width at half-maximum (FWHM) pulse duration is $2\sqrt{2\log{2}}\sigma$.
For the coherent pump processes necessary for quantum state interconversion, the following connection is often required:
\begin{equation}
\int_{0}^{4\sigma}\alpha\Omega_0\mathrm{exp}[-\frac{(t-2\sigma)^{2}}{2\sigma^{2}}]dt={\pi},
\end{equation}
where $\alpha=\sqrt{3}$ for steps $(1)$ and $(5)$ of $\mathcal{I}$, $\alpha=\sqrt{2}$ for step $(5)$ of $\mathcal{II}$, and $\alpha=1$ for the rest coherent driving processes. Note that we have applied the constraint that the pulse area of $\pi$ must be attained within a limited period of $4\sigma$, thus the associated parameter $\sigma$ can be determined as follows
\begin{equation}
\sigma=\frac{\sqrt{\pi}}{\sqrt{2}\alpha\Omega_0\mathrm{Erf}[{\sqrt{2}}]}.
\end{equation}

In Figs.~\ref{error}(c) and \ref{error}(d),
we examine the effects of distance errors and timing errors on the dissipative quantum state interconversion process under the same number of cycles with $\Omega_0=2\pi\times0.072~\rm MHz$.
In comparison to the rectangular-pulse (RP) schemes, the Gaussian-pulse (GP) schemes can further increase the resilience of the distance fluctuations between atoms, and even shortens the preparation time for the $W$ state and the GHZ state by $34.5~\mu\rm s$ and $43.3~\mu\rm s$, respectively.

\subsection{Comparison with unitary dynamics-based protocols}\label{VI0}
\begin{figure}
\centering\scalebox{0.44}{\includegraphics{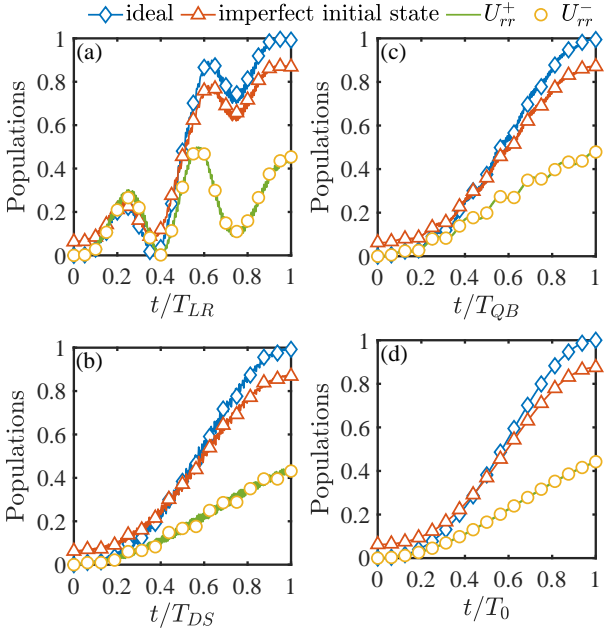}}
\caption{\label{compare} Time evolution of population for GHZ state in Ref.~\cite{16} (a),  Ref.~\cite{17} (b), Ref.~\cite{69} (c), and  Ref.~\cite{67} (d) under different conditions, where $T_{LR}=17.83~\mu{\rm s}$, $T_{DS}=3.1018~\mu{\rm s}$, $T_{QB}=2.3322~\mu{\rm s}$, and  $T_{0}=T_{DS}$.}
\end{figure}

As stated in the introduction, there are currently four strategies for the interconversion of GHZ and $W$ states in neutral-atom systems based on unitary dynamics. These include the Lewis-Riesenfeld invariants-based scheme \cite{16}, the dynamic-symmetry-based approach \cite{17}, the quantum-brachistochrone approach \cite{69}, and the simplified dynamic protocol without a strong off-resonant laser field \cite{67}, respectively.
Taking the $W$-to-GHZ state conversion $\mathcal{II}$ as an example, we reproduce the conversion processes of the above four schemes (diamond) in Fig.~\ref{compare} with parameters $T_{LR}=17.83~\mu{\rm s}$, $T_{DS}=3.1018~\mu{\rm s}$, $T_{QB}=2.3322~\mu{\rm s}$, and  $T_{0}=T_{DS}$. The final forms of GHZ states prepared by these schemes are
$|{\rm GHZ_{(a)}}\rangle=(|000\rangle+\exp[-i\Sigma_{i=1}^3(\delta_i+\Delta_i)T_{LR}]|111\rangle)/\sqrt{2}$,
$|{\rm GHZ_{(b)}}\rangle=(|000\rangle+\exp\{i\pi/2-i3[U_{rr}-2\Omega_{r0}^2(\Delta_0-U_{rr})/(\Delta^2_0-2\Delta_0U_{rr})]T_{DS}\}|111\rangle)/\sqrt{2}$,
$|{\rm GHZ_{(c)}}\rangle=(|000\rangle+\exp\{-i\pi/2-i3[U_{rr}-2\Omega_{r0}^2(\Delta_0-U_{rr})/(\Delta^2_0-2\Delta_0U_{rr})]T_{QB}\}|111\rangle)/\sqrt{2}$,
and $|{\rm GHZ_{(d)}}\rangle=(|000\rangle+\exp(i\pi/2-3iU_{rr}T_{0})|111\rangle)/\sqrt{2}$, where the definition of each parameter can be found in the corresponding literature. We note that there is always a time-dependent relative phase between the components of the GHZ state, which will inevitably cause the final state to be susceptible to fluctuations in system parameters. Specifically, we investigate
the effects of imperfect initial states ($\rho_{0}^{ip}=7/8|W_{0}\rangle\langle W_{0}|+1/8|000\rangle\langle 000|$) and fluctuations of the Rydberg interaction ($U_{rr}^{\pm}=U_{rr}\pm0.1U_{rr}$ with $U_{rr}$ the desired vdW potential) on these schemes and plot the corresponding temporal evolution of populations in Fig.~\ref{compare}. Compared to the ideal case of each scheme, these two factors will cause the population of the target state to deviate significantly from unity. The effect of the fluctuations of the Rydberg interaction is especially significant, resulting in a population of less than 50\% for the GHZ state within the parameter range we consider. In Table \ref{tab5}, we also list in detail the populations of the target states based on the RP scheme and the GP scheme under different conditions and compare them with the previous unitary dynamics-based schemes.
It is worth reemphasizing that the fidelity of the GHZ state, created through dissipation, can be enhanced by increasing the number of cycles---a possibility unavailable in schemes relying on unitary dynamics. This aspect holds utmost significance. Additionally, in Appendix \ref{D}, we explore the impact of optical frequency fluctuations of the resonant driving field $\Omega_2$ on the dissipative quantum state interconversion, and the aforementioned characteristic remains observable in this investigation.

\begin{table}
	\caption{Comparison of our scheme to previous unitary dynamics-based schemes under different conditions in the process of $W$-to-GHZ state conversion.}\label{tab5}
	\centering
	\begin{tabular}{cccc}
		\hline
		\hline
		{Protocols} & {Ideal case}& {Imperfect initial state}  & {Imperfect $U_{rr}$}
		\\
		\hline
        RP scheme    &{$99.32\%$}  &{$99.33\%$}   &$97.60\%~(93.97\%)$\\
		GP scheme    &{$99.55\%$}  &{$99.56\%$}   &$97.92\%~(98.27\%)$\\
		Ref.~\cite{16}     &{$99.19\%$}   &{$86.85\%$}  &{$45.12\%~(45.50\%)$}\\
		Ref.~\cite{17}     &{$99.24\%$}   &{$86.83\%$}  &{$44.05\%~(42.91\%)$}\\
		Ref.~\cite{69}     &{$99.44\%$}    &{$87.05\%$} &{$47.67\%~(48.11\%)$}\\
		Ref.~\cite{67}      &{$100\%$}    &{$87.50\%$} &{$44.24\%~(44.24\%)$}\\
		\hline
		\hline
	\end{tabular}
\end{table}

\section{Conclusions and outlook}\label{V}
In this work, we have presented a dissipative technique for attaining the interconversion between the GHZ and $W$ states in Rydberg atom arrays. The use of periodic pump and dissipation, on the one hand, allows us to optimize the design of coherent and incoherent operations without requiring a delicate balance between them, and, on the other hand, can adjust the steady-state form of the system in a fixed physical system by simply varying the driving fields, which is the key to achieving the bidirectional conversion between GHZ state and $W$ state.
Numerically, using the Fock-Liouville space, it is proved that the target state in each conversion process is the unique steady state in the Floquet-Lindblad framework.
In conjunction with the experimentally accessible parameters at the current neutral-atom platform, the effects of laser phase noise, interatomic distance fluctuations, and timing errors (intensity fluctuations of the Rabi frequency) on the efficiency of quantum state interconversion are studied, and it is found that our current scheme has strong robustness to these factors. Compared to earlier quantum state interconversion systems based on unitary dynamics, our work can provide a higher-fidelity and more robust interconversion for two different kinds of genuine tripartite entanglement from the experimental point of view.

Our protocol can be directly generalized to the parallel creation of distinct forms of tripartite entangled states in a two-dimensional array of Rydberg atoms \cite{PhysRevLett.123.170503}. These entangled states distributed in each unit cell can be utilized as resource states to build fusion-based quantum computation, which relies on finite-sized entangled states and destructive entangling measurements to achieve fault-tolerant quantum computing \cite{bartolucci2023fusion}. Moreover, the products of three GHZ states can form a gauge logic bit for the Bacon-Shor code, which has been proven to be  beneficial for fault-tolerant control of an error-corrected
qubit \cite{egan2021fault}.
Thus, our scheme has the potential to be utilized in these error-tolerant quantum computations involving Rydberg atoms.

Several additional improvements are possible.
We have shown, for instance, that the introduction of Gaussian pulses can stabilize the system into a more resilient target state in less time than the use of rectangular pulses. Hence, we predict that new types of pulses generated by the optimal quantum control algorithm would be able to significantly reduce the execution time of the system while maintaining the high fidelity of the target state \cite{PhysRevLett.106.190501, PhysRevA.84.022326, PhysRevA.92.062343}. In addition, although intuitive, the Floquet-Lindblad scenario is only sufficient but not necessary for the dissipative interconversion of quantum states in our physical system. If we can formulate the necessary coherent and incoherent operations in the form of the Kraus operators and map them to the quantum-circuit model, we may achieve an efficient quantum state interconversion protocol by optimizing the overall quantum circuit \cite{PhysRevLett.101.060401, PhysRevA.98.032309}.

We anticipate that our method will provide a fresh perspective on the interconversion of GHZ and $W$ states in neutral-atom systems, and enable the preparation of more intricate multipartite and high-dimensional entangled states, which are pertinent to quantum information processing. We look forward to its experimental realization shortly.

\section*{ACKNOWLEDGMENTS}

This work is supported by the National Natural Science Foundation of China (NSFC) under Grant No. 12174048. W.L. acknowledges support from the EPSRC through Grant No.\ EP/W015641/1.

\appendix
\section{Effective energy level structure obtained by two-photon process}\label{A}
\begin{figure}
\centering\scalebox{0.38}{\includegraphics{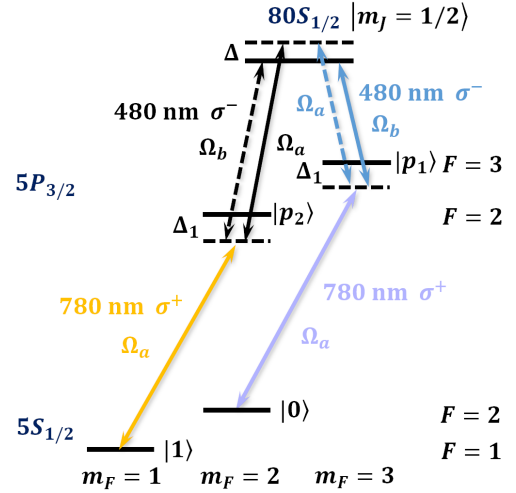}}
\caption{\label{fulllevel} Relevant levels of $^{87}$Rb for realizing the effective single-photon transitions of Fig.~\ref{pulse} with a series of two-photon excitations.}
\end{figure}

The single-photon transitions shown in Fig.~\ref{pulse} are carried out in a genuine atomic system by a series of two-photon processes, as detailed in Fig.~\ref{fulllevel}.
For the actualization of the laser-driven three-level atomic system shown in Fig.~\ref{pulse}(b1), a $\sigma^+$-polarized $780$-${\rm nm}$ laser beam of Rabi frequency $\Omega_a$ may be exploited to couple the $|0\rangle\leftrightarrow|p_1\rangle=|5P_{3/2}, F=3,m_F=3\rangle$ transition with a red detuning $\Delta_1$, whereas a  $\sigma^-$-polarized $480$-${\rm nm}$ laser beam with a Rabi frequency $\Omega_b$ is tuned $\Delta_1$ to the blue of the $|p_1\rangle\leftrightarrow|r\rangle$ transition. The mechanism described above enables the resonant transition between states $|0\rangle$ and $|r\rangle$.
Similarly, the dispersive coupling between states $|1\rangle$ and $|r\rangle$ necessitates the use of two extra laser fields, where $|1\rangle\leftrightarrow|p_2\rangle=|5P_{3/2},F=2,m_F=2\rangle$ transition is driven by a $\sigma^+$-polarized $780$-${\rm nm}$ laser beam with a Rabi frequency $\Omega_a$ and a red detuning $\Delta_1$, while $|p_2\rangle\leftrightarrow|r\rangle$ transition is achieved via another $\sigma^-$-polarized $480$-${\rm nm}$ laser beam with a Rabi frequency $\Omega_a$, blue detuned by $\Delta_1+\Delta$. After adiabatically eliminating the intermediate
state $|p_{1(2)}\rangle$ in the limit of large detuning $\Omega_{a(b)}\ll\{\Delta_1,\Delta_1+\Delta\}$ and neglecting the ac Stark shifts,
we can obtain the effective laser-atom interaction Hamiltonian as
\begin{equation}\label{fu1}
H_{(b1)}^{\rm eff}=\frac{\Omega_a\Omega_b}{4\Delta_1}|0\rangle\langle r|
+\frac{(2\Delta_1+\Delta)\Omega_a^2}{8\Delta_1(\Delta_1+\Delta)}|1\rangle\langle r|e^{i\Delta t}+{\rm H.c.}
\end{equation}

To realize the laser-driven two-level atomic system depicted in Fig.~\ref{pulse}(c1), we only need to introduce an additional  $\sigma^-$ polarized $480$-${\rm nm}$ laser beam of Rabi frequency $\Omega_a$ that is tuned $\Delta_1+\Delta$ to the blue of the $|p_1\rangle\leftrightarrow|r\rangle$ transition to induce the off-resonant transition from state $|0\rangle$ to state $|r\rangle$, on the basis of retaining the resonant $|0\rangle\leftrightarrow|r\rangle$ transition. In the large-detuning regime $\Omega_{a(b)}\ll\{\Delta_1,\Delta_1+\Delta\}$, we have
\begin{equation}\label{fu2}
H_{(c1)}^{\rm eff}=\frac{\Omega_a\Omega_b}{4\Delta_1}|0\rangle\langle r|
+\frac{(2\Delta_1+\Delta)\Omega_a^2}{8\Delta_1(\Delta_1+\Delta)}|0\rangle\langle r|e^{i\Delta t}+{\rm H.c.}
\end{equation}
The corresponding relationship between the effective coupling strengths of Eqs.~(\ref{fu1}) and (\ref{fu2}) and the Rabi frequencies used in the text reads
\begin{equation}
\frac{(2\Delta_1+\Delta)\Omega_a^2}{8\Delta_1(\Delta_1+\Delta)}\rightleftharpoons\frac{\Omega_1}{2},\  \
\frac{\Omega_a\Omega_b}{4\Delta_1}\rightleftharpoons\frac{\Omega_2}{2}.
\end{equation}
Fixing $\Delta_1=5\Delta=2\pi\times1000~{\rm MHz}$, $\Omega_a=2\pi\times93.42~{\rm MHz}$, and $\Omega_b=2\pi\times0.856~{\rm MHz}$ enables us to obtain the desired parameters $\Omega_1=2\pi\times4~{\rm MHz}$, $\Omega_2=2\pi\times0.04~{\rm MHz}$, and $\Delta=2\pi\times200~{\rm MHz}$.
It is important to note that the driving mode may be readily extended to other kinds required by the coherent pump operations during the quantum state interconversion.

\section{The mechanism of the entangled pump and selective excitation interpreted by quantum Zeno dynamics}\label{B}
In the interaction picture, the entangled pump Hamiltonian of the system in  Fig.~\ref{pulse}(b1) is given by
\begin{eqnarray}\label{fullmurp0}
H_{(b1)}&=&\sum_{j=1}^{3}\frac{\Omega_{1}}{2}e^{-i\Delta t}|r_j\rangle\langle 1_j|+\frac{\Omega_{2}}{2}|r_j\rangle\langle 0_j|+{\rm H.c.}\nonumber\\&&+\sum_{j<k}U_{rr}|r_j\rangle\langle r_j|\otimes|r_k\rangle\langle r_k|.
\end{eqnarray}
Here, we choose the quantized $z$-axis perpendicular to the two-dimensional plane of the Rydberg atom array, so that the influence of the phases caused by the laser wave vectors can be ignored.
We use the formula $i\dot{U}_{0}^{\dag}U_{0}+U_{0}^{\dag}H_{(b1)}U_{0}$ with $U_{0}=\textrm{exp}\{-it\sum_{j<k}U_{rr}|r_j\rangle\langle r_j|\otimes|r_k\rangle\langle r_k|\}$ to move to a rotating frame and obtain the transformed Hamiltonian as
\begin{equation}\label{IIee7}
H_{(b1)}=H_{\Omega_{1}}+H_{\Omega_{2}}+H_{D},
\end{equation}
where we have assumed $U_{rr}=\Delta$ and each term of Eq.~(\ref{IIee7})  is represented by
\begin{subequations}\label{333}
\begin{eqnarray}
H_{\Omega_{1}}&=&\sum^{3}_{j=1}\frac{\Omega_{1}}{2}\sigma^{r1}_{j}(P^{r}_{j+1}P^{0}_{j+2}+P^{r}_{j+1}P^{1}_{j+2}
+P^{0}_{j+1}P^{r}_{j+2}\nonumber\\&&+P^{1}_{j+1}P^{r}_{j+2})+\mathrm{H.c.},\label{IIe8aa}\\ H_{\Omega_{2}}&=&\sum^{3}_{j=1}\frac{\Omega_{2}}{2}\sigma^{r0}_{j}(P^{0}_{j+1}P^{0}_{j+2}+P^{0}_{j+1}P^{1}_{j+2}+P^{1}_{j+1}P^{0}_{j+2}\nonumber\\&&+P^{1}_{j+1}P^{1}_{j+2})+\mathrm{H.c.},\label{IIe8b}\\
H_{D}&=&\sum^{3}_{j=1}\frac{\Omega_{1}}{2}e^{-i\Delta t}\sigma^{r1}_{j}(P^{0}_{j+1}P^{0}_{j+2}+P^{0}_{j+1}P^{1}_{j+2}\nonumber\\&&+P^{1}_{j+1}P^{0}_{j+2}+P^{1}_{j+1}P^{1}_{j+2})+\frac{\Omega_{1}}{2}e^{i\Delta t}\sigma^{r1}_{j}P^{r}_{j+1}P^{r}_{j+2}\nonumber\\&&+\frac{\Omega_{2}}{2}e^{i\Delta t}\sigma^{r0}_{j}(P^{r}_{j+1}P^{0}_{j+2}+P^{r}_{j+1}P^{1}_{j+2}+P^{0}_{j+1}P^{r}_{j+2}\nonumber\\&&+P^{1}_{j+1}P^{r}_{j+2})+\frac{\Omega_{2}}{2}e^{2i\Delta t}\sigma^{r0}_{j}P^{r}_{j+1}P^{r}_{j+2}+\mathrm{H.c.},\label{IIe8c}\nonumber\\
\end{eqnarray}
\end{subequations}
with $P_{j}^{a}=|a_j\rangle\langle a_j|$ (here $P_{j-3}^{a}=P_{j}^{a}$ for $j>3$) and $\sigma^{ab}_{j}=|a_j\rangle\langle b_j|$  being the projection operator and the transition operator of $j$th atom, respectively.
The contribution of Eq.~(\ref{IIe8c}) is reduced to the Stark shifts of the atomic energy levels under the condition that the detuning $\Delta$ is substantially larger than the coupling strength $\Omega_{1}$$(\Omega_{2})$, which may be counteracted by introducing additional fields and auxiliary energy levels.

The remaining parts may be rewritten as $H^{'}_{(b1)}=\Omega_{2}/2(K{\cal H}_{\Omega_{1}}+{\cal H}_{\Omega_{2}})$, where $K=\Omega_1/\Omega_2$ and ${\cal H}_{\Omega_{1}}$ $({\cal H}_{\Omega_{2}})$ becomes the dimensionless interaction Hamiltonian between classical fields and atoms. In the limit of $K\rightarrow\infty~(\Omega_{1}\gg\Omega_{2})$, it can be shown that the unitary operator is regulated by $U^{'}_{(b1)}(t)\sim \mathrm{exp}[-i\Omega_{2}/2(\sum_{n}K\varepsilon_{n}{\cal P}_nt+{\cal H}_{z}t)]$, where ${\cal P}_n$ is the eigenprojection of ${\cal H}_{\Omega_{1}}$ belonging to the eigenvalue $\varepsilon_{n}$, and ${\cal H}_{z}=\sum_{n}{\mathscr {\cal P}_n}{\cal H}_{\Omega_{2}}{{\cal P}_n}$ is referred to as the Zeno Hamiltonian~\cite{PhysRevLett.85.1762, PhysRevLett.89.080401, Shao_2010epl,
signoles2014confined,barontini2015deterministic,bretheau2015quantum}. The Hamiltonian of the system may further be reduced to $H_{\rm EP0}=\Omega_{2}/2{{\cal P}_0}{\cal H}_{\Omega_{2}}{{\cal P}_0}={{\cal P}_0}H_{\Omega_{2}}{{\cal P}_0}$ if the subspace of interest is in the projection operator with eigenvalue 0.
Therefore, the effective Hamiltonian of Eq.~(\ref{fullmurp0}) is obtained as
\begin{eqnarray}\label{murp0}
H_{\rm EP0}&=&\frac{\sqrt{3}\Omega_{2}}{2}|D_{0}\rangle\langle 000|-\frac{\sqrt{3}\Omega_{2}}{4}(|\psi_{0}^{2}\rangle+|\psi_{0}^{3}\rangle)\langle W_{0}^{\prime}|\nonumber\\&&-\frac{\Omega_{2}}{4}(2|\psi_{0}^{1}\rangle+|\psi_{0}^{2}\rangle-|\psi_{0}^{3}\rangle)\langle W_{0}^{\prime\prime}|+\textrm{H.c.}
\end{eqnarray}

\begin{figure}
\centering\scalebox{0.57}{\includegraphics{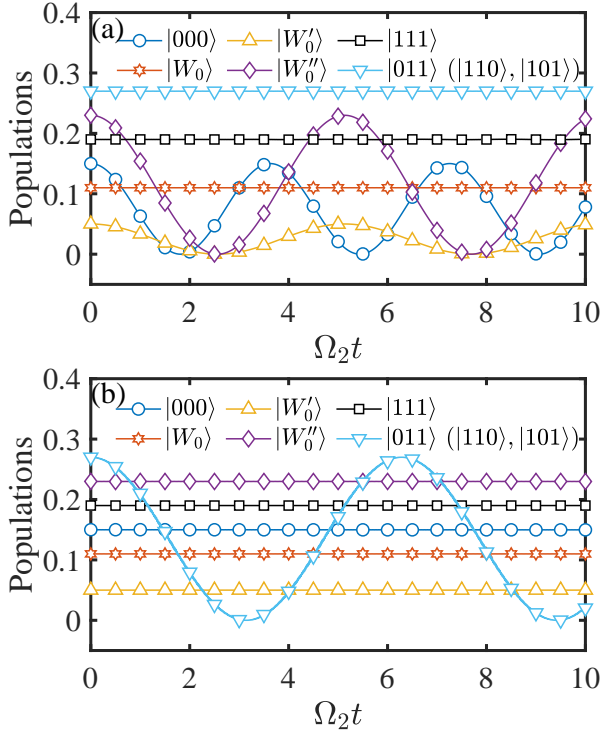}}
\caption{\label{a3}(a) The populations as functions of $\Omega_{2}t$ under the full Hamiltonian Eq.~(\ref{fullmurp0}) (solid lines) and the effective Hamiltonian Eq.~(\ref{murp0}) (scattered points) for the case of EP0. (b) The populations as functions of $\Omega_{2}t$ under the full Hamiltonian Eq.~(\ref{IIe1}) and the effective Hamiltonian Eq.~(\ref{urp0}) for the case of SE0. For the two situations described above, the initial state is set to be  $\rho_{0}=0.19|111\rangle\langle 111|+0.05|W_{0}^{\prime}\rangle\langle W_{0}^{\prime}|+0.11|W_{0}\rangle\langle W_{0}|+0.23|W_{0}^{\prime\prime}\rangle\langle W_{0}^{\prime\prime}|+0.15|000\rangle\langle 000|+0.27|011\rangle\langle 011|$, and other parameters are selected as $\Delta/\Omega_{1}=50$ and $\Omega_{1}/\Omega_{2}=100$.}
\end{figure}

For the interaction model illustrated in Fig.~\ref{pulse}(c1), the Hamiltonian in the interaction picture is given by
\begin{eqnarray}\label{IIe1}
H_{(c1)}&=&\sum_{j=1}^{3}\frac{\Omega_{1}}{2}e^{-i\Delta t}|r_j\rangle\langle 0_j|+\frac{\Omega_{2}}{2}|r_j\rangle\langle 0_j|+{\rm H.c.}\nonumber\\&&+\sum_{j<k}U_{rr}|r_j\rangle\langle r_j|\otimes|r_k\rangle\langle r_k|.
\end{eqnarray}
By using a similar process for derivation from Eq.~(\ref{fullmurp0}) to Eq.~(\ref{333}), i.e., rotating the Hamiltonian first, then removing the Stark shifts, finally separating the remaining resonant interaction strong coupling
\begin{eqnarray}
H_{\Omega_{1}}^{'}&=&\sum^{3}_{j=1}\frac{\Omega_{1}}{2}\sigma^{r0}_{j}(P^{r}_{j+1}P^{0}_{j+2}+P^{r}_{j+1}P^{1}_{j+2}+P^{0}_{j+1}P^{r}_{j+2}\nonumber\\&&+P^{1}_{j+1}P^{r}_{j+2})+\mathrm{H.c.},\label{IIe8a}
\end{eqnarray}
 and weak coupling
\begin{eqnarray}
H_{\Omega_{2}}^{'}&=&\sum^{3}_{j=1}\frac{\Omega_{2}}{2}\sigma^{r0}_{j}(P^{0}_{j+1}P^{0}_{j+2}+P^{0}_{j+1}P^{1}_{j+2}+P^{1}_{j+1}P^{0}_{j+2}\nonumber\\&&+P^{1}_{j+1}P^{1}_{j+2})+\mathrm{H.c.},\label{IIe8bb}
\end{eqnarray}
the effective Hamiltonian of Eq.~(\ref{IIe1}) under the condition $U_{rr}=\Delta\gg\Omega_1\gg\Omega_2$ is obtained as
\begin{eqnarray}\label{urp0}
H_{\textrm{SE0}}&=&{{\cal P}^{'}_0}H_{\Omega_{2}}^{'}{{\cal P}^{'}_0}\nonumber\\
&=&\frac{\Omega_{2}}{2}(|r11\rangle\langle 011|+|1r1\rangle\langle 101|+|11r\rangle\langle 110|)+\mathrm{H.c.},\nonumber\\
\end{eqnarray}
where ${{\cal P}^{'}_0}$ is the eigenprojection of ${H}^{'}_{\Omega_{1}}$ belonging to the eigenvalue $0$.

The entangled pump and selective excitation constitute the coherent and incoherent operations necessary for the dissipative interconversion between the GHZ and $W$ states.
To check the accuracy of the preceding derivation, we utilize the full Hamiltonians of Eqs.~(\ref{fullmurp0}) and (\ref{IIe1}) (solid lines) to simulate the dynamic evolution process of an initial state $\rho_{0}=0.19|111\rangle\langle 111|+0.05|W_{0}^{\prime}\rangle\langle W_{0}^{\prime}|+0.11|W_{0}\rangle\langle W_{0}|+0.23|W_{0}^{\prime\prime}\rangle\langle W_{0}^{\prime\prime}|+0.15|000\rangle\langle 000|+0.27|011\rangle\langle 011|$ in Fig.~\ref{a3} (the states in parentheses can be used to replace the component of the mixed state in the density operator, but the evolution remains unchanged) and compare these results to the effective Hamiltonians of Eqs.~(\ref{murp0}) and (\ref{urp0}) (scattered points), respectively. It can be seen that within a given range of parameters $\Delta/\Omega_{1}=50$ and $\Omega_{1}/\Omega_{2}=100$, the effective Hamiltonians properly describe the evolution characteristics of the system, i.e., for the laser-atom interaction models shown in Fig.~\ref{pulse}, EP0 can ensure the stability of the $|W_{0}\rangle=(|100\rangle+|010\rangle+|001\rangle)/\sqrt{3}$ state, but SE0 can only drive a system with only one atom in state $|0\rangle$.

\section{The engineered spontaneous emission}\label{C}
\begin{figure}
\centering\scalebox{0.38}{\includegraphics{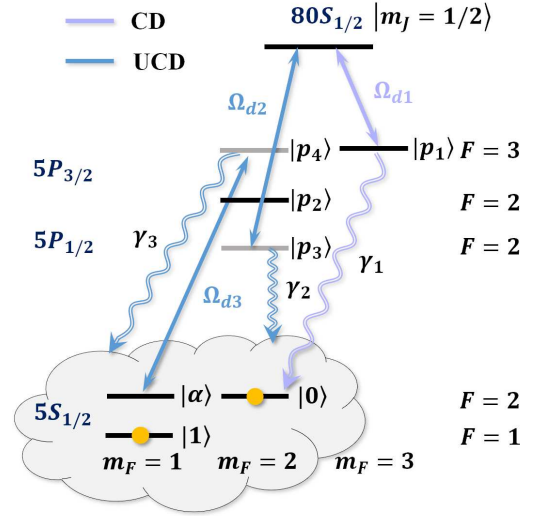}}
\caption{\label{decay}The schematic diagram of the engineered spontaneous emission. The rapid decay of atoms can be realized by choosing different paths. For the dissipation channel CD, a resonant laser (Rabi frequency $\Omega_{d_1}$) is employed to couple the Rydberg state $|r\rangle$ to a short-lived state $|p_1\rangle$ (lifetime $1/\gamma_1\simeq 26.26$~{$\rm ns$}) which only decays into $|0\rangle$. For the dissipation channel UCD, a resonant laser (Rabi frequency $\Omega_{d_2}$) is used to couple the Rydberg state $|r\rangle$ to another short-lived state $|p_{3}\rangle=|5P_{1/2}, F=2, m_F=2\rangle$ (lifetime $1/\gamma_{2}\simeq 27.68$~{$\rm ns$}) which then decays into $|0\rangle$, $|1\rangle$, and $|\alpha\rangle=|5S_{1/2}, F=2,m_F=1\rangle$ with probabilities $1/3$, $1/2$ and $1/6$, respectively. The noncomputational state $|\alpha\rangle$ caused by spontaneous emission of $|p_3\rangle$ is coupled resonantly with a short-lived state $|p_{4}\rangle=|5P_{3/2},F=3,m_F=2\rangle$ (lifetime $1/\mathrm{\gamma}_3=1/\gamma_1$) through another laser (Rabi frequency $\Omega_{d_{3}}$) and then $|P_4\rangle$ decays into $|\alpha\rangle$ and $|0\rangle$ with probabilities $2/3$ and $1/3$, respectively.}
\end{figure}

The schematic diagram of the engineered spontaneous emission is demonstrated in Fig.~\ref{decay}.
For the realization of the dissipation channel CD, we use a $\sigma^-$-polarized 480~{$\rm nm$}  laser with a Rabi frequency $\Omega_{d_1}$ to couple
the transition between the Rydberg state $|r\rangle$
and the intermediate state $|p_{1}\rangle$ which can only decay into the ground state $|0\rangle$ at rate $\mathrm{\gamma}_{1}\simeq 2\pi\times6.06$~{$\rm MHz$} owning to the transition selection rule. The evolution of the
density matrix $\rho$ for a single atom is described by the Lindblad master
equation
\begin{equation}\label{1}
\dot{\rho}=-i[H_{d_{1}},\rho]+\gamma_{1}\mathcal{D}[|0\rangle\langle p_{1}|]\rho,
\end{equation}
where $H_{d_{1}}=\Omega_{d_{1}}/2|r\rangle\langle p_{1}|+\mathrm{H.c.}$. In the limit of large decay rate $\gamma_{1}\gg\Omega_{d_{1}}$, the short-lived state $|p_{1}\rangle$ can be adiabatically eliminated and the engineered decay rate $\Gamma_{1}=\Omega_{d_{1}}^{2}/\mathrm{\gamma_{1}}$ can be obtained. Therefore, Eq.~(\ref{1}) can be rewritten as
\begin{equation}\label{2}
\dot{\rho}=\Gamma_{1}\mathcal{D}[|0\rangle\langle r|]\rho,
\end{equation}
which can be solved analytically. For an atom that initially in
the state $|r\rangle$, the probability of ever finding it in the ground
state $|0\rangle$ is
\begin{equation}
\rho_{00}(t)=1-e^{-\Gamma_{1}t}.
\end{equation}
\begin{figure}
\centering\scalebox{0.44}{\includegraphics{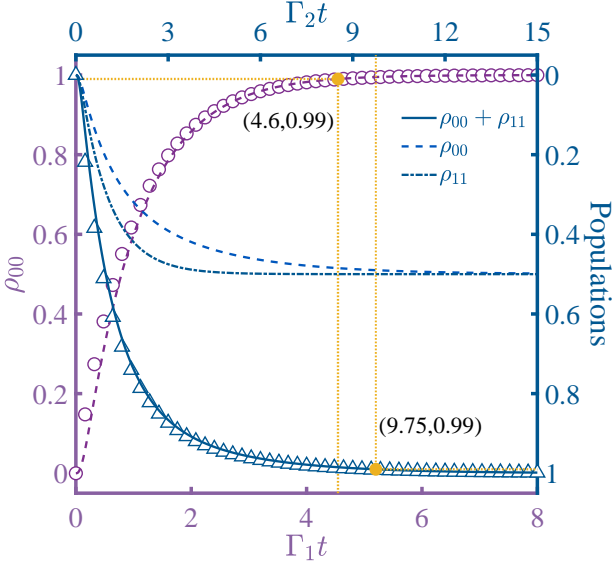}}
\caption{\label{ndecay}The performance of the dissipation channels CD and UCD via the method of engineered spontaneous emission. The solid lines and scatter points represent values obtained from the full master equation and the effective master equation, respectively. The initial state is chosen as $|r\rangle$ and other parameters $\Omega_{d_{i}}/\gamma_{i}=0.2~(i=1,2,3)$. For the UCD, the system can be stabilized in the mixed state of $|0\rangle$ and $|1\rangle$ with equal probabilities.}
\end{figure}

To achieve the dissipation channel UCD, we employ a $\pi$-polarized $474$~{$\rm nm$} laser of Rabi frequency $\Omega_{d_2}$ to couple the same Rydberg state $|r\rangle$ to a short-lived state $|p_{3}\rangle=|5P_{1/2}, F=2, m_F=2\rangle$ (decay rate $\gamma_2\simeq2\pi\times5.75$~{$\rm MHz$}) which decays into ground states $|0\rangle$, $|1\rangle$, and a noncomputational basis state $|\alpha\rangle=|5S_{1/2},F=2,m_F=1\rangle$ with probabilities $1/3$, $1/2$, and $1/6$, respectively.

To recycle the population of the noncomputational basis state $|\alpha\rangle$, a $\sigma^+$-polarized $780$~{$\rm nm$} laser of Rabi frequency $\Omega_{d_3}$ is introduced to pump atom to the state $|p_{4}\rangle=|5P_{3/2}, F=3, m_F=2\rangle$ (decay rate $\gamma_3=\gamma_1$),  and the probabilities of spontaneous
emission of this state to states $|\alpha\rangle$ and $|0\rangle$ are $2/3$ and $1/3$,
respectively. In this case, the master equation of the model
reads
\begin{equation}\label{3}
\dot{\rho}=-i[H_{d_{2}},\rho]+\sum_{k=2,3}\sum_{l=0,1,\alpha}\gamma_{k}^{l}\mathcal{D}[|l\rangle\langle p_{k+1}|]\rho,
\end{equation}
where $H_{d_{2}}=\Omega_{d_{2}}/2|r\rangle\langle p_{3}|+\Omega_{d_{3}}/2|\alpha\rangle\langle p_{4}|+\mathrm{H.c.}$, and the branching
ratio of decay rates $\gamma_{2}^{0}=\gamma_{2}/3$, $\gamma_{2}^{1}=\gamma_{2}/2$, $\gamma_{2}^{\alpha}=\gamma_{2}/6$, $\gamma_{3}^{0}=\gamma_{3}/3$, $\gamma_{3}^{1}=0$, and $\gamma_{3}^{\alpha}=2\gamma_{3}/3$. Similarly, the short-lived states $|p_{3}\rangle$ and $|p_{4}\rangle$ can be adiabatically eliminated in the limit of large decay rate $\gamma_{i}\gg\Omega_{d_{i}}$ $(i=2,3)$, then Eq.~(\ref{3}) can be rewritten as
\begin{equation}\label{4}
\dot{\rho}=\sum_{l=0,1,\alpha}\Gamma_{2}^{l}\mathcal{D}[|l\rangle\langle r|]\rho+\Gamma_{3}^{l}\mathcal{D}[|l\rangle\langle \alpha|]\rho,
\end{equation}
where the branching ratio of the engineered decay rates $\Gamma_{2}^{0}=\Gamma_{2}/3$, $\Gamma_{2}^{1}=\Gamma_{2}/2$, $\Gamma_{2}^{\alpha}=\Gamma_{2}/6$, $\Gamma_{3}^{0}=\Gamma_{3}/3$, $\Gamma_{3}^{1}=0$, and $\Gamma_{3}^{\alpha}=2\Gamma_{3}/3$ with $\Gamma_i=\Omega^2_{di}/\gamma_i$.
We can directly obtain the temporal
evolution of the total populations of the computational states $|0\rangle$ and $|1\rangle$
\begin{equation}
\rho_{00}(t)+\rho_{11}(t)=1-e^{-\Gamma_{2}t}-\frac{\Gamma_{2}}{6\Gamma_{2}-
2\Gamma_{3}}(e^{-\frac{\Gamma_{3}t}{3}}-e^{-\Gamma_{2}t}),
\end{equation}
from the initial condition $\rho(0)=|r\rangle\langle r|$.
In Fig.~\ref{ndecay}, we compare the full process (solid lines) with the effective process (scattered points) under the parameters  $\Omega_{d_{i}}/\gamma_{i}=0.2$ $(i=1,2,3)$ and $\Gamma_3=\Gamma_2\gamma_3/\gamma_2$.
It is discovered that with
just $\tau_{1}=4.6/\Gamma_1\simeq3~\mu\mathrm{s}$ and $\tau_{2}=9.75/\Gamma_2\simeq6.7~\mu\mathrm{s}$, respectively, the probability
of atomic radiation to the ground states may approach 99$\%$ for
the dissipative process CD and UCD, and then the evolution of the system Eq.~(\ref{4}) can be equivalently replaced by
\begin{equation}\label{eff}
\dot{\rho}=\frac{\Gamma_{2}}{2}\mathcal{D}[|0\rangle\langle r|]\rho+\frac{\Gamma_{2}}{2}\mathcal{D}[|1\rangle\langle r|]\rho.
\end{equation}
\section{ The fluctuations of optical frequency of the
resonant driving field}\label{D}
\begin{figure}
\centering\scalebox{0.44}{\includegraphics{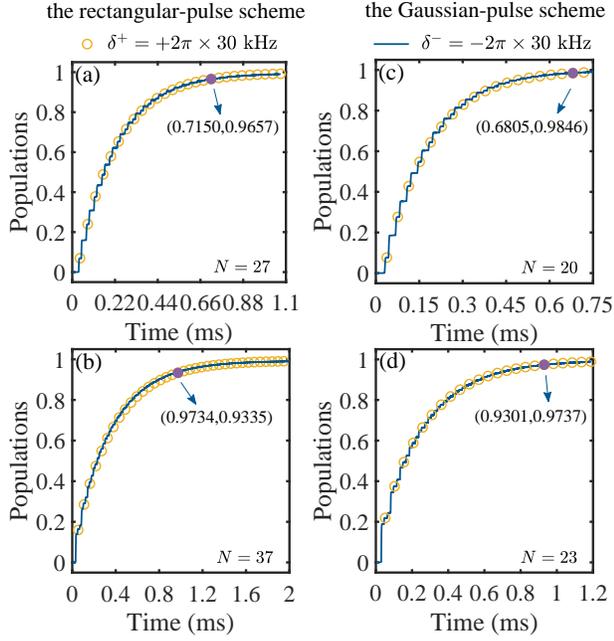}}
\caption{\label{delta} The influence of the frequency fluctuations of the resonant driving field $\Omega_2$ on the dissipative interconversion between GHZ and $W$ states. (a) and (b) represent the conversion from GHZ state to $W$ state and its reverse process, respectively, using a rectangular pulse.  (c) and (d) illustrate the corresponding conversions using a Gaussian pulse.}
\end{figure}

In the Rydberg-atom system experiments, the typical sources of noise also include fluctuations in the laser optical frequency and Rabi frequency.
In Sec.~\ref{dt}, the effects of variations in interatomic spacing and timing errors during the coherent operations have been discussed in detail. These two effects also reflect, respectively, the optical frequency fluctuations of the strong driving field $\Omega_1$ and the Rabi frequency fluctuations of the weak driving field $\Omega_2$.

The variations in interatomic spacing will lead to $U_{rr}\neq\Delta$ in realistic situations.
In addition to the variations in interatomic distance, this effect can be attributed to the optical frequency fluctuations of the strong driving field.
 In each coherent operation, the timing errors may lead to the pulse area $\int_0^{t_i+\delta t}\Omega_2dt$, which can be reformulated as $\int_0^{t_i}(\Omega_2+\delta\Omega_2)dt$ with $\int_0^{t_i}\delta\Omega_2dt=\int_0^{\delta t}\Omega_2dt$, thus the influence of the fluctuations of the Rabi frequency can be equivalently reflected as the errors in the time selection, as we claimed before.
It should be noted that we do not need to simultaneously consider the Rabi frequency fluctuations of two laser fields ($\Omega_1$ and $\Omega_2$) because our scheme requires only $\Omega_2\ll\Omega_1$. Consequently, the additional noise that we need to consider should be the fluctuations of the optical frequency of the weak laser field $\Omega_2$.

Here we introduce a detuning parameter $\delta$, which describes the discrepancy between the optical frequency of the weak laser field and the associated transition frequency of the atom.
In Fig.~\ref{delta}, we explore the impact of optical frequency fluctuations of the resonant driving field $\Omega_2$ on the dissipative quantum state interconversion.
Whether $\delta$ is positive (red-detuned) or negative (blue-detuned) $2\pi\times30~{\rm kHz}$, the effect on the quantum state interconversion scheme is negative, and the populations of target states are inferior to that of the resonant driving (as indicated by the solid purple circles for the red-detuning laser). Fortunately, we can increase the number of cycles to further boost the scheme's performance. For the population of the final target state to be close to $99\%$, the number of cycles for executing the GHZ-to-$W$ state conversion and its reverse process using the rectangular pulse must be increased to $27$ and $37$, respectively, whereas the Gaussian pulse-based scheme only needs to increase the number of cycles to $20$ and $23$, respectively. These findings suggest that the dissipative quantum state interconversion scheme is preferable to the unitary dynamics-based scheme. In addition, it demonstrates that time-dependent modulation can bring more optimized results to our scheme.

\bibliography{ref}

\end{document}